\documentclass[a4paper,11pt]{article}
\usepackage{amssymb,graphicx,amsmath}
\usepackage{contract}

\makeatletter 
\renewcommand\section{\@startsection{section}{1}{\z@}              {-3.25ex\@plus -1ex \@minus -.2ex}                                    {1.5ex \@plus .2ex}                                    {\reset@font\Large\bfseries}}
\renewcommand\subsection{\@startsection{subsection}{2}{\z@}                                    {3.25ex \@plus 1ex \@minus.2ex}                                    {-1em}                                    {\reset@font\large\bfseries}}
\@addtoreset{equation}{section} \makeatother
\renewcommand{\theequation}{\thesection.\arabic{equation}}
\input{tcilatex}

\newcommand{\bc}{\begin{center}}
\newcommand{\ec}{\end{center}}
\def\ba#1{\begin{array}{#1}\displaystyle}
\newcommand{\ea}{\end{array}}
\newcommand{\z}{\\[2mm] \displaystyle}

\newcommand{\beq}{\begin{equation}}
\newcommand{\eeq}{\end{equation}}
\newcommand{\beqa}{\begin{eqnarray}}
\newcommand{\eeqa}{\end{eqnarray}}
\newcommand{\no}{\nonumber}
\newcommand{\n}{\nonumber\\}
\newcommand{\bi}{\begin{itemize}}
\newcommand{\ei}{\end{itemize}}

\def\lt#1{\left#1}
\def\rt#1{\right#1}
\def\t#1{\tilde{#1}}
\def\h#1{\hat{#1}}
\def\b#1{\bar{#1}}
\def\frc#1#2{\frac{#1}{#2}}

\newcommand{\bra}{\langle}
\newcommand{\ket}{\rangle}
\newcommand{\Z}{{\mathbb{Z}}}

\newcommand{\ep}{\epsilon}
\newcommand{\varep}{\varepsilon}

\newcommand{\shift}{{\rm shift}}

\begin{document}

\setcounter{page}{0} \topmargin0pt \oddsidemargin0mm \renewcommand{%
\thefootnote}{\fnsymbol{footnote}} \newpage \setcounter{page}{0}
\begin{titlepage}

\vspace{0.2cm}
\begin{center}
{\Large {\bf Bi-partite entanglement entropy in massive QFT with a boundary: the Ising model}}

\vspace{0.8cm} {\large  \text{Olalla
A.~Castro-Alvaredo$^{\bullet}$ and Benjamin Doyon$^{\circ}$}}

\vspace{0.2cm}{$^{\bullet}$  Centre for Mathematical Science, City University London, \\
Northampton Square, London EC1V 0HB, UK} \\
\vspace{0.2cm} {$^{\circ}$  Department of Mathematical Sciences,
Durham University \\ South Road, Durham DH1 3LE, UK }
\end{center}
\vspace{1cm}

\noindent In this paper we give an exact infinite-series expression for the
bi-partite entanglement entropy of the quantum Ising model both with
a boundary magnetic field and in infinite volume. This generalizes and extends previous
results involving the present authors for the bi-partite
entanglement entropy of integrable quantum field theories, which
exploited the generalization of the form factor program to
branch-point twist fields. In the boundary case, we isolate in a
universal way the part of the entanglement entropy which is related
to the boundary entropy introduced by Affleck and Ludwig, and explain how this relation
should hold in more general QFT models. We
provide several consistency checks for the validity of our form
factor results, notably, the identification of the leading
ultraviolet behaviour both of the entanglement entropy and of the
two-point function of twist fields in the bulk theory, to a great
degree of precision by including up to 500 form factor
contributions.

\vfill{ \hspace*{-9mm}
\begin{tabular}{l}
\rule{6 cm}{0.05 mm}\\
$^\bullet \text{o.castro-alvaredo@city.ac.uk}$\\
$^\circ \text{benjamin.doyon@durham.ac.uk}$\\
\end{tabular}}

\renewcommand{\thefootnote}{\arabic{footnote}}
\setcounter{footnote}{0}
\end{titlepage}
\newpage
\section{Introduction}

Entanglement is a fundamental characteristic of quantum systems,
which has great importance, for instance, in the context of
quantum computing. Any measure of entanglement is likely to give a
good description of the quantum nature of a ground state. Many
measures of entanglement have been devised, see e.g.
\cite{bennet}-\cite{Verstraete}. However, perhaps due to its
geometric character which makes it more theoretically tractable in
systems with many degrees of freedom, the entanglement entropy
\cite{bennet} has attracted great interest in theoretical physics.
The entanglement entropy can also be argued to be a better measure
of fundamental properties of the ground state than correlation
functions, as it is not associated to a particular observable, but
rather to a sector of mutually local observables. In this paper we
continue the research initiated in \cite{entropy,other,next} on
the bi-partite entanglement entropy in one-dimensional quantum
models with many local degrees of freedom, focusing on the effect
of boundaries.

The entanglement entropy is a measure of quantum entanglement
between the degrees of freedom of two regions, $A$ and its
complement $\bar{A}$, in some quantum state $|\psi\rangle$.
Consider a quantum system, with Hilbert space ${\cal H} = {\cal
H}_A\otimes{\cal H}_{\b{A}}$, in a pure state $|\psi\rangle$. The
bipartite entanglement entropy $S_A$ is the von Neumann entropy
associated to the reduced density matrix $\rho_A$ of the subsystem
$A$,
\begin{equation}
    \rho_A = \text{Tr}_{{\cal H}_{\b{A}}}(|\psi\rangle\langle \psi|)\,,
\end{equation}
given by
\begin{equation}\label{toi}
   S_{A}=-\text{Tr}_{\mathcal{H}_A} (\rho_A \log \rho_A)= - \lim_{n \rightarrow 1}\frac{d}{d n} \text{Tr}_{\mathcal{H}_A}
   (\rho_A^n).
\end{equation}
The expression with the $n$-limit on the right-hand side is often referred to as
the replica-trick.

Let us consider now the scaling limit of the quantum system, describing the universal part of the behaviour
near a quantum critical point. It is obtained by
approaching the critical point while letting the length of the
region $A$ go to infinity in a fixed proportion with the
correlation lengths. The result is a quantum field
theory (QFT) model, which we will take throughout to possess
($1+1$-dimensional) Poincar\'e invariance. In what follows, we consider only the case where $A$ is a connected region.

In the QFT context, the expression on the right-hand side of the
second equation of (\ref{toi}) can be understood, for $n$ a
natural number, as a normalised partition function for the model
on a Riemann surface with two branch points, with $n$ sheets
cyclicly connected, or on a surface with two conical singularities
of angles $2\pi n$ \cite{bombelli,CallanW94,Calabrese:2005in} (see
also the explanations in \cite{entropy,other}). There is only one
way of associating such branch points to well-defined local QFT
fields. This was first done in \cite{entropy} in the case
\emph{without boundaries} (with $A$ a region in the bulk). There,
it was shown how to relate the entanglement entropy in
two-dimensional QFT with a two-point function of certain local
fields defined in a model consisting of $n$ copies of the original
model, called {\em branch-point twist fields} (section
\ref{general}):
\begin{equation}
   S_A^{\rm bulk}(r)=
   - \lim_{n \rightarrow 1}\frac{d}{d n} {\cal Z}_n \varepsilon^{4\Delta_n}\langle 0| \t{{\cal T}}(x_1) {\cal
    T}(x_2)|0\rangle.\label{def}
\end{equation}
Here $A$ has length $r=|x_1-x_2|$ and $\langle
0|\cdots|0\rangle$ denote correlation functions in the $n$-copy model; the state
$|0\rangle$ is the vacuum state of the latter. The derivative with respect to $n$
involves an appropriate analytic continuation in $n$ of the correlation function, which is
assumed to be in correspondence with the conical-singularity interpretation. We will not
discuss further in the present paper the subtleties and assumptions involved in this analytic continuation
-- see the discussion in \cite{other} for more details. The constant ${\cal Z}_n$, with ${\cal Z}_1=1$, is an
$n$-dependent non-universal constant, $\varepsilon$ is a short-distance cut-off
which is chosen so that $d {\cal Z}_n/dn=0$ and, finally, $\Delta_n$ is
the conformal dimension of the counter parts of the fields
$\mathcal{T}, \tilde{\mathcal{T}}$ in the underlying $n$-copy conformal
field theory,
\begin{equation}\label{dn}
    \Delta_n=\frac{c}{24}\left(n-\frac{1}{n}\right),
\end{equation}
which can be obtained by CFT arguments \cite{Calabrese:2005in,entropy} and where $c$ is the
central charge.

In \cite{entropy,other}, the two-point function $\langle 0|
\t{\cal T}(x_1) {\cal T}(x_2)|0\rangle$ was studied at large
distances $r$ for all 1+1-dimensional integrable QFTs on the line.
The main feature of these models is that there is no particle
production in any scattering process and the scattering ($S$)
matrix factorizes into products of 2-particle $S$-matrices which
can be calculated exactly (for reviews see e.g.
\cite{Karowski:1978eg}-\cite{Dorey:1996gd}). Taking the $S$-matrix
as input it is then possible to compute the matrix elements of
local operators (also called form factors). This is done by
solving a set of consistency equations \cite{KW,Smirnovbook}. This
is known as the form factor bootstrap program for integrable QFTs.
In \cite{entropy,other}, this program was used and generalised in
order to compute the two-particle approximation of $\langle 0|
\t{\cal T}(x_1) {\cal T}(x_2)|0\rangle$, and to obtain the leading
correction to saturation of the entanglement entropy. This leading
correction was observed to be very universal, as it is independent
of the scattering matrix. In fact, by similar techniques, this
universal correction was also observed to hold outside of
integrability \cite{next}.

In this paper we generalise the construction above in order to
include the presence of one integrable boundary. The study of
integrable QFTs with boundaries has attracted a lot of attention
in the last two decades (see e.g.
\cite{Cherednik:1985vs}-\cite{Bowcock:1995vp}). The present work,
will make extensive use of the results of S.~Ghoshal and A.~B.
Zamolodchikov \cite{Ghoshal:1993tm}, particularly the explicit
realization of the boundary which they proposed. Their work
provided also a detailed study of the Ising model, for which the
integrable boundary conditions were classified and the
corresponding reflection amplitudes computed. These reflection
amplitudes will provide a crucial input for our entropy
computation.

Let us therefore consider a family of integrable boundaries
parametrised by the dimensionless constant \beq\label{defkappa}
    \kappa=1-h^2/(2m)\in (-\infty,1)~,
\eeq related to a uniform magnetic field $h$ affecting the
boundary Ising spins. We study the entanglement entropy between a
region $A$ that extends from the boundary to a distance $r$, and
the rest. For $\kappa \leq 0$, we obtain the full large-distance
series expansion; the result depends on the reflection matrix at
all orders. It turns out that some of the techniques necessary to
obtain this result are also useful in the bulk case, so that as a
by-product, we obtain the equivalent expansion in the Ising model
without boundaries, extending our previous work \cite{entropy}.
We note that this extension to higher particle
contributions involves subtleties that were not present in the
two-particle case.

We then evaluate from the form factor expansion the exact universal
constant $V(\kappa)$ that relates the large-distance value of the entanglement entropy to the
short-distance logarithmic behaviour in the {\em boundary} case, defined by
\beq\label{shla}
    S_A^{\rm boundary}(r) = \lt\{ \ba{ll} \frc{c}6 \log(2r/\varepsilon) + V(\kappa) + o(1) & \varepsilon \ll r \ll m^{-1} \z
    -\frc{c}6 \log(\varepsilon m) + \frc{U}2 + O((rm)^{-\infty})& r\gg m^{-1} \ea\rt.
\eeq
where $m$ is the mass of the particle of the Ising model and $c=1/2$ is its central charge
(and $o(1)$ is in terms of the small combination $rm$). Note that the corrections to the large-distance saturation
are exponential. The constant $U$ is the universal saturation constant that occurs in the bulk case,
calculated in the Ising model in \cite{entropy}:
\beq\label{shlabu}
    S_A^{\rm bulk}(r) = \lt\{ \ba{ll} \frc{c}3 \log(r/\varepsilon) + o(1) & \varepsilon \ll r \ll m^{-1} \z
    -\frc{c}3 \log(\varepsilon m) + U + O((rm)^{-\infty})& r\gg m^{-1} \ea\rt.
\eeq Equations (\ref{shlabu}) and (\ref{shla}) provide universal
definitions for $U$ and $V(\kappa)$. Note that $\varep$ here is in
general a different short-distance cut-off in the bulk and
boundary cases. The choice of these definitions will become clear
later.

Our main findings are 1) the observation of non-monotonicity of
the entanglement entropy in the boundary Ising model for boundary
magnetic field lower than a critical value, and 2) the relation
between the constant $V(\kappa)$ and the boundary degeneracy $g$ of
Affleck and Ludwig \cite{Affleck:1991tk}. Exact
re-summations of form factors and CFT arguments strongly suggest
the following result: \beq\label{resVkappa}
    V(\kappa) = \lt\{\ba{ll} \log\sqrt{2} & (\kappa>-\infty) \\ 0 & (\kappa=-\infty). \ea \rt.
\eeq In more general cases of massive QFT, for $h$ associated to a relevant boundary perturbation, we argue that \beq\label{resVkappagen}
    V(\kappa) = s - \log {\cal C},
\eeq where $s=\log g$ is the boundary entropy in the UV ($\kappa>-\infty$) or infrared ($\kappa=-\infty$),
and ${\cal C}^2$ is the fraction of the massive ground state degeneracy that is broken by the field $h$.
This provides a way to extract the boundary
entropy solely from universal entanglement entropies.

The paper is organised as follows: in section \ref{general} we
introduce the bi-partite entanglement entropy of a general massive
boundary QFT. We establish a relationship between this quantity
and the derivative at $n=1$ of the boundary one-point function of
a twist field associated to the QFT constructed as $n$
non-interacting copies of the original model, similarly as
discussed for the bulk case in \cite{entropy,other,next}. We provide a CFT analysis
of the constant $V(\kappa)$ introduced above. We then specialize these results
to the case of integrable models, introducing the form factor expansion
for the boundary entanglement entropy. Finally we introduce the Ising
model, for which we review the different kinds of integrable
boundary conditions and the associated reflection matrices
obtained in \cite{Ghoshal:1993tm}. In section 3 we start by reviewing the form factor
approach for branch-point twist fields and give a closed
expression for all non-vanishing form factors of the Ising model.
We proceed to a detailed analysis of the boundary entanglement entropy in the Ising model, giving a closed expression
for all form factor contributions. The derivation of these formulae involves a
complicated analytic continuation on the variable $n$, as the
derivative is taken, followed by the $n\rightarrow 1$ limit.  In
section 4 a similar analysis is performed for the bulk theory,
extending the results of \cite{entropy}. In section 5 we provide a
analytical and numerical study of the ultraviolet behaviour
of the entanglement entropy of the bulk and boundary Ising model, and evaluate $V(\kappa)$.
In section 6 we provide a discussion of the main results, and in section 7
we summarize our main conclusions and open problems. We
provide three appendices: In appendix A we give a proof of several
formulae which we have used for the computation of the
entanglement entropy.  In appendix B we provide alternative
formulae for the individual form factor contributions to the bulk
and boundary entanglement entropy which are more suitable for
numerical computations. In appendix C we provide a detailed
analysis of the UV behaviour of the two-point function of the
twist fields in the bulk Ising model and extract the coefficient
of the logarithmic term with great precision.

\section{Entanglement entropy of two-dimensional QFTs with boundaries} \label{general}

\subsection{General considerations}\label{gencons} \indent \\

\noindent We consider here a general massive boundary QFT with one
of the particle masses being $m$, characterising the scale of all
other masses, and a boundary parameter $h$, associated to a
relevant perturbation of a conformal boundary condition. More
precisely, we may write the action as \beq
     S = S_{\rm bulk}(m) + h \int dt\, \phi(t),\label{action}
\eeq where $\phi(t)$ is a boundary field with dimension less than
1 and $t$ is time. For later convenience, we will use
$\kappa=1-h^2/(2m) \in (-\infty,1)$.

The main idea of \cite{entropy} in order to evaluate the
entanglement entropy was to use $n$ independent copies of the
original model, which preserves integrability. The introduction of
extra copies allows for the presence of a $\mathbb{Z}_n$ symmetry,
associated to which the branch-point twist fields $\mathcal{T}(x)$
and $\tilde{\mathcal{T}}(x)$ can be defined. It is these twist
fields that implement the non-trivial connection between sheets in
the Riemann surface with branch points, used to calculate the
entanglement entropy. Below we recall some of their main
properties, but for more details about branch-point twist fields
and the construction in the bulk case, see the discussions in
\cite{entropy,other}.

The branch-point twist fields can be characterized as
follows: let $\Psi_{1}, \ldots, \Psi_{n}$ be any fields
belonging to each copy of the original model\footnote{The vector space of fields
of the multi-copy model contains the $n$-fold tensor product of the vector space of fields of the
original model. Fields belonging to a copy $i$ have the identity field ${\bf 1}$ for all factors corresponding
to copies $j\neq i$.}.
Then the equal time ($x^0=y^0$) exchange relations between
$\mathcal{T}(x)$ and $\Psi_{1}(y), \ldots, \Psi_{n}(y)$ can be written in
the following form\footnote{Here we employ the standard notation
in Minkowski space-time: $x^{\nu}$ with $\nu=0,1$, with $x^{0}$
being the time coordinate and $x^{1}$ being the position
coordinate.}:
\begin{eqnarray}
    \Psi_{i}(y)\mathcal{T}(x) &=& \mathcal{T}(x) \Psi_{i+1}(y) \qquad x^{1}> y^{1},
    \nonumber\\
    \Psi_{i}(y)\mathcal{T}(x) &=& \mathcal{T}(x) \Psi_{i}(y) \qquad x^{1}< y^{1}, \label{cr}
\end{eqnarray}
for $i=1,\ldots, n$ and where we identify the indices $n+i \equiv
i$ and similarly
\begin{eqnarray}
    \Psi_{i}(y)\tilde{\mathcal{T}}(x) &=& \tilde{ \mathcal{T}}(x) \Psi_{i-1}(y) \qquad x^{1}> y^{1},
    \nonumber\\
    \Psi_{i}(y)\tilde{\mathcal{T}}(x) &=& \tilde{ \mathcal{T}}(x) \Psi_{i}(y) \qquad x^{1}< y^{1}, \label{cr2}
\end{eqnarray}
that is $\mathcal{T}=\tilde{\mathcal{T}}^\dagger$. These exchange relations indicate that a branch cut
originates from branch-point twist fields in (Euclidean) correlation functions with insertion of fields $\Psi_i$.
The definition of the branch-point twist field is completed
by saying that they are spinless, invariant under all symmetries of the original model, and of lowest dimension,
and by specifying the CFT normalisation $\t{\mathcal{T}}(x) {\cal T}(0) \sim |x|^{-4\Delta_n}$ as $|x|\to0^+$ (space-like),
with the conformal dimension given in (\ref{dn}). Note that as $n\to1$, ${\cal T}\to {\bf 1}$.

We now consider the presence of a boundary. We will
place the boundary at the origin of space $x^1=0$ and therefore
have a QFT defined on the (positive) half-line. In order to make a clear connection with
the bulk results, we first consider
the entanglement entropy $S(r_1,r_2)$ in the case where the
region $A$ is a bulk region, extending from $r_1>0$ to $r_2>r_1$, region $\b{A}$ being the rest, composed of two
disconnected components, from $0$ to $r_1$ and from $r_2$ to $\infty$. Arguments entirely similar to those
of \cite{entropy,other} show that the trace in (\ref{toi}) becomes a normalised correlation function
of twist fields as in (\ref{def}), but with the ground state of the model on the half-line, $|0\ket_B$. Note in particular that the
boundary condition is $\Z_n$-invariant, so that the branch cuts originating from the twist fields can
be deformed through the boundary.

Then, the entanglement entropy is given by:
\begin{equation}\label{be}
   S_A(r_1,r_2)=
   - \lim_{n \rightarrow 1}\frac{d}{d n} {\cal Z}_n \varepsilon^{4\Delta_n}
    {}_B\langle 0|{\tilde{{\cal T}}}(r_2) {{\cal
    T}}(r_1)|0\rangle_B
\end{equation}
Again, the non-universal constant ${\cal Z}_n$ satisfies ${\cal Z}_1=1$ and $d {\cal Z}_n/dn=0$,
$\varepsilon$ is a short-distance cut-off, and $\Delta_n$ is given in (\ref{dn}).
In this equation, we have a vacuum correlation function in the model
on the half-line, the twist-fields being at positions $r_1$ and $r_2$ on the half-line.

There are two ways to obtain the entanglement entropy $S_A$ for a
region $A$ starting from the boundary and ending at $r$. First, we
could consider the limit $r_1\to0$ in $S(r_1,r)$, making the bulk
region $[r_1,r]$ approach the boundary. As the twist field at
$r_1$ approaches the boundary, the correlation function diverges,
because the presence of the boundary changes the regularisation
necessary around the branch point. A way to evaluate the
divergency is to use boundary conformal field theory, which applies in massive models
when a local field is near to a boundary. It tells
us that for small $r_1$ there is a power law determined by the conformal dimension of ${\cal T}$:
${}_B\bra0| \cdots {\cal T}(r_1)|0\ket_B \propto  r_1^{-2\Delta_n}$ \cite{CardyLewellen}. We
may then define ${\cal T}(0)|0\ket_B$ as $\lim_{r_1\to0}
r_1^{2\Delta_n} {\cal T}(r_1)|0\ket_B$. This appropriately
regularised operator ${\cal T}(0)$ is simply proportional to the
unitary operator performing a $\Z_n$ transformation, since its
branch cut, through which $\Z_n$ transformations are performed, now
extends through the whole space. But since $|0\ket_B$ is invariant
under such a transformation, we find, with appropriate choice of
proportionality constants,
\begin{equation}\label{be2}
   S_A^{\rm boundary}(r)=
   - \lim_{n \rightarrow 1}\frac{d}{d n} {\cal Z}_n \varepsilon^{2\Delta_n}{}_B\langle 0|{{\cal T}}(r)|0\rangle_B,
\end{equation}
where we have changed the power of $\varep$ in order to keep a scaling dimension of 0 (essentially,
this accounts for the change of regularisation necessary around the branch points). We used the fact that the entropy is real in order
to change $\t{\cal T}\to {\cal T}$ by complex conjugation.

Second, we may take the limit $r_2\to\infty$ in $S(r,r_2)$. Then, the two-point function in (\ref{be}) reduces to its disconnected part:
\beq
    {}_B\langle 0|{\tilde{{\cal T}}}(r_2) {{\cal
    T}}(r)|0\rangle_B \sim {}_B\langle 0|{\tilde{{\cal T}}}(\infty)|0\ket_B {}_B\bra0| {{\cal
    T}}(r)|0\rangle_B~.
\eeq
In the first factor, the twist field does not feel the presence of the boundary, hence this expectation value can be replaced
by its expectation value in the model without boundary, $\bra 0| {\cal T}|0 \ket$. Dividing out this factor and using the appropriate
branch-point regularisation, we find again (\ref{be2}).


Some of these considerations are made clearer by using
crossing in order to implement the boundary as a state:
\beq
    {}_B\langle 0|{\tilde{{\cal T}}}(r_2) {{\cal
    T}}(r_1)|0\rangle_B =  \langle 0|{\tilde{{\cal T}}}(r_2) {{\cal T}}(r_1)|B\rangle.
\eeq The boundary state $|B\ket$ is in the past at time 0 in the
Hilbert space of the model {\em on the full line}, and the twist
fields are placed at {\em imaginary times} $r_1$ and $r_2$. More
precisely, the boundary state is the $n$-fold tensor product of
single-copy boundary states. The state $\bra0|$ is the ground
state of the $n$-copy model on the line, corresponding to
asymptotic conditions at positive infinite times. No factor occurs
in using crossing symmetry since the branch-point twist fields are
spinless. The normalisation of the boundary state $|B\ket$ is such
that $\bra0|B\ket = 1$. Using crossing, we get
\begin{equation}\label{be3}
   S_A^{\rm boundary}(r)=
   - \lim_{n \rightarrow 1}\frac{d}{d n} {\cal Z}_n \varepsilon^{2\Delta_n}\langle 0| {{\cal T}}(r)|B\rangle.
\end{equation}
It is this approach that we will be exploiting in this paper.

\subsection{Large and short distance behaviour}\label{shortdistbeh} \indent \\

\noindent In the calculations above, we did not keep track of the
normalisation constants occurring in reaching (\ref{be3}). These
correspond to an additive constant to the entanglement entropy,
which is not universal. However, as mentioned in the introduction,
the difference between constants occurring at large and short
distances is universal. We now show that the expression
(\ref{be3}) gives the choice of large distance behaviour
(\ref{shla}). When $r$ is very large, we may use a decomposition
of the identity, between the twist field and the boundary state,
in energy eigenstates (see (\ref{sk})). The leading term comes
from the ground state, giving \beq
    S_A^{\rm boundary}(r) \sim - \lim_{n \rightarrow 1}\frac{d}{d n} {\cal Z}_n \varepsilon^{2\Delta_n}\langle 0| {{\cal T}}|0\ket \quad (rm\to\infty)
\eeq
using $\bra0|B\rangle=1$. From \cite{entropy}, this is exactly $-(c/6)\log (\varep m) + U/2$ as defined in (\ref{shlabu}),
hence this shows (\ref{shla}).

We may give a CFT expression for the short-distance constant
$V(\kappa)$. From the general formula for one-point functions in a
boundary CFT with boundary state $|B\ket$ \cite{CardyLewellen}, we
have \beq\label{sd}
    \bra 0 |{\cal T}(r)|B\ket \sim \bra {\cal T}|B\ket^{CFT} (2r)^{-2\Delta_n}
\eeq
as $rm\to0$. Here, $|{\cal T}\ket$ is the normalised highest weight state corresponding to the primary field ${\cal T}$ in CFT.
In the cases of finite perturbing boundary parameter $h$, the state $|B\ket^{CFT}$
is the UV limit $h\to0$ of the boundary state $|B\ket$. In the case
where $h\to\infty$ before $rm\to0$, the state $|B\ket^{CFT}$ is the corresponding IR limit.
Hence from (\ref{be3}), we have
\beq
    S_A^{\rm boundary}(r) \sim \frc{c}6 \log(2r/\varepsilon) - \lim_{n\to1} \frc{d}{dn} \bra {\cal T}|B\ket^{CFT} + o(1) \quad (rm\to0)
\eeq
which gives
\beq\label{Vkapgen}
    V(\kappa) = - \lim_{n\to1} \frc{d}{dn} \bra {\cal T}|B\ket^{CFT}
\eeq
where we recall that $\bra 0 |B\ket^{CFT} = 1$, with $\bra 0 |0\ket=1$ and $\bra {\cal T}|{\cal T}\ket = 1$.
We see here that $V(\kappa)$ (with the definition (\ref{defkappa}) for $\kappa$)
takes only two values, one for $\kappa$ finite (UV) and one for $\kappa=-\infty$ (IR), in agreement with (\ref{resVkappa}). This is the
main conclusion that we can derive from (\ref{Vkapgen}), as it is a non-trivial
matter to evaluate this expression, and there may be subtleties associated to massive ground state degeneracies (see section \ref{discuss}).

Note finally that we can write $V(\kappa)$ solely in terms of entanglement entropies:
\beq
    V(\kappa) = \lim_{\eta\to0} \lt(S_A^{\rm boundary}(\eta) - S_A^{\rm boundary}(\eta^{-1})
    - \frc12 S_A^{\rm bulk}(\eta) +\frc12 S_A^{\rm bulk}(\eta^{-1}) - \frc{c}6 \log
    2\rt),
\eeq
where boundary and bulk entanglement entropies may be evaluated in different cut-off schemes.

\subsection{Integrable models and large-distance expansion}\indent \\

\noindent For simplicity, let us consider a model of integrable
QFT with a single particle spectrum and no bound states. The
asymptotic states forming a basis of the Hilbert space are
characterised by a number of particles $k$ and by the rapidities
$\theta_j$ of these particles. As usual in QFT, two bases can be
defined, representing particles coming in ($in$-states) and
particles going out ($out$-states). In the $n$-copy model, we will
denote the $in$-states by
$|\theta_1,\ldots,\theta_k\rangle_{a_1,\ldots,a_k}$ with
$\theta_1>\ldots>\theta_k$, where $a_j$ are the copy labels; the
vacuum will be denoted $|0\ket$. The scattering matrix describes
the linear relations between the two bases. The two-particle
scattering matrix of the $n$-copy model, that depends only on the
rapidity difference $\theta=\theta_1-\theta_2$ by relativistic
invariance, is given by
\begin{equation}\label{s}
    S_{ab}(\theta)=S(\theta)^{\delta_{ab}}, \qquad \text{for}
    \qquad a,b=1 \ldots n,
\end{equation}
with $S(\theta)$ being the scattering matrix of the original
model and $a,\,b$ the copy labels of the two particles.
That is, the different copies of the model do not
interact with each other.

As mentioned above, the state $|B\rangle$ is just a tensor product
of boundary states in the individual copies. In integrable models,
these have an explicit expression as the famous boundary state
introduced by Ghoshal and Zamolodchikov \cite{Ghoshal:1993tm}. In the case where no boundary
bound state can form, we have
\begin{equation}\label{bs}
|B\rangle =\exp\left(\frac{1}{4
\pi}\sum_{j=1}^{n}\int_{-\infty}^{\infty}
R\left(\frac{i\pi}{2}-\theta\right) Z_{j}(-\theta)Z_j
(\theta)\right)|0\rangle.
\end{equation}
The function $R(\theta)$ is the boundary reflection matrix of the
integrable QFT and $Z_j(\theta)$ are the Faddeev-Zamolodchikov
operators, which provide a generalization of the
creation-anhilation operators for integrable QFTs with non-trivial
interactions \cite{ZZ,Faddeev:1980zy}. Their main properties are
\beqa
    && Z_{a_1}(\theta_1) \cdots Z_{a_k}(\theta_k) |0\rangle = |\theta_1,\ldots,\theta_k\rangle_{a_1,\ldots,a_k} \mbox{ for }
    \theta_1>\ldots>\theta_k \n
&&Z_{a}(\theta_1)Z_b(\theta_2) = S_{ab}(\theta_1-\theta_2) Z_b(\theta_2)Z_a(\theta_1). \no
\eeqa
The tensor-product form of the boundary state indicates that particles living in different copies
of the theory do not interact through the presence of the
boundary.

One can now expand in (\ref{be3}) the boundary operator defined above to obtain a large-$r$ expansion:
\begin{equation}\label{sk}
 \langle 0| {\mathcal{T}}(r)|B \rangle = \bra{\cal T}\ket \sum_{\ell=0}^{\infty}
    f_{\ell}(2rm,\kappa),
\end{equation}
where
\begin{eqnarray}\label{1p}
    \bra{\cal T}\ket f_{\ell}(t,\kappa)&=&\frac{1}{\ell!
    (4\pi)^\ell}\sum_{j_1,j_2,\ldots,j_\ell=1}^{n} \left[\prod_{r=1}^{\ell} \int_{-\infty}^{\infty} d \theta_{r}e^{-t \cosh \theta_{r}}
    R\left(\frac{i\pi}{2}-\theta_{r}\right)\right]\nonumber
    \\
    & \times & F_{2\ell}^{\mathcal{T}|j_1 j_1 j_2 j_2\ldots j_\ell j_\ell}
    (-\theta_{1},\theta_{1},\ldots,-\theta_{\ell},\theta_{\ell}),
\end{eqnarray}
and
\begin{equation}\label{ff}
   F_{\ell}^{\mathcal{T}|j_1 \ldots j_\ell}(\theta_1, \ldots,
   \theta_\ell)=\langle 0| \mathcal{T}(0) Z_{j_1}(\theta_1) \cdots Z_{j_\ell}(\theta_\ell)|0\rangle
\end{equation}
are the $\ell$-particle form factors of the operator
${\mathcal{T}}$ in the $n$-copy model. For example:
\begin{eqnarray}
  f_0(t,\kappa)
    &=&  1, \\
  f_1(t,\kappa) &=&\frac{n}{4
    \pi \bra{\cal T}\ket}\int_{-\infty}^{\infty}d\theta\,
    R\left(\frac{i\pi}{2}-\theta\right)
    F_{2}^{\mathcal{T}|11}(-\theta,\theta)e^{-t \cosh \theta} ,\label{2p}
\end{eqnarray}
and so on. This is a useful expansion, because the form factors of
branch-point twist fields can be obtained exactly, in principle,
by solving a set of consistency equations, first given in
\cite{entropy}. We will calculate these form factors in the next
section for the Ising model.

Employing (\ref{be3}) and (\ref{dn}), the entanglement entropy is
given by
\begin{equation}
    S_A(rm) =-\frac{c}{6}\log(\varepsilon m)+ \frc{U}2 + \sum_{\ell=1}^{\infty}
    s_{\ell}(2rm,\kappa).\label{ent}
\end{equation}
with
\begin{equation}\label{sell}
   s_{\ell}(2rm,\kappa) = \left.-\frac{d f_{\ell}(2rm,\kappa)}{dn}\right|_{n=1},
\end{equation}
and where
\begin{equation}
    U =-\frac{d}{dn} [m^{-4 \Delta_n}
    \langle\mathcal{T}\rangle^2]_{n=1}
\end{equation}
is a universal constant, that relates the large-distance
saturation to the short-distance logarithmic behaviour in the bulk
case. In obtaining the expression (\ref{ent}) for the entanglement
entropy, we used the fact that ${\cal T}$ becomes the identity
operator at $n=1$, so that $\langle\mathcal{T}\rangle_{n=1}=1$ and
all its form factors with one or more particles vanish.

It is this form factor expansion that we will use in the sections
that follow  in order to evaluate the entanglement entropy in the
boundary Ising model and in particular the constant $V(\kappa)$.

\subsection{Integrable boundaries in the Ising model} \label{sectRIsing}\indent \\

\noindent The massive Ising model is characterised by the fact
that the two-particle scattering matrix, as defined above, is
$S(\theta)= -1$ (in the single-copy model): the particles are free
Majorana Fermions. The central charge in (\ref{ent}) is 1/2 and
the constant
\begin{equation}\label{u}
   U=-0.131984...
\end{equation}
was obtained in \cite{entropy,Jin}. Let us now recall the types of
integrable boundary conditions that have been found for the Ising
model. A family that was studied in much detail in
\cite{Ghoshal:1993tm} is that corresponding to the presence of a
magnetic field that couples to the Ising spin field ($\phi(t)$ in
the action (\ref{action})) on the boundary. To be precise, the
spin field is the order parameter, hence we are looking at the
scaling limit of the Ising spin chain in a transverse magnetic
field whose magnitude is slightly below its critical value, and
with a parallel magnetic field on the boundary. The corresponding
boundary reflection matrix is given by
\begin{equation}\label{r}
    R(\theta)=-i \tanh\frac{1}{2}\left(\theta-\frac{i \pi}{2}\right)\frac{\kappa-i \sinh\theta}{\kappa+ i
    \sinh\theta},
\end{equation}
which includes, for special values of the parameter $\kappa$,
the following physically different types of integrable boundary conditions,
\begin{itemize}
    \item \emph{{Free boundary condition}}: $\kappa=1$,
    \item \emph{{Fixed boundary condition}}: $\kappa = -\infty$,
    \item \emph{{Magnetic boundary conditions}} (interpolating between the previous
    two): $\kappa=1-\frac{h^2}{2m}$, where $h$ is a boundary
    magnetic field $0 < h < \infty$. The free boundary condition
    would then correspond to $h=0$ whereas the fixed boundary
    condition is equivalent to having a infinitely large magnetic
    field fixed at the boundary.
\end{itemize}
Boundary corrections to the expectation values of the energy and
disorder field in the Ising theory were computed using this reflection matrix in
\cite{Konik:1995ws}.

In the cases where $\kappa>0$, the reflection matrix has a pole on
the imaginary line on the physical sheet, $0<{\rm Im}(\theta)\leq
i\pi/2$. This implies that the boundary state expression
(\ref{bs}) is not correct. A modified expression exists
\cite{Ghoshal:1993tm}, but for simplicity, in the present paper we
will not analyse this case. Hence, throughout we will consider
$\kappa\leq 0$ (except in sub-section \ref{twofour}). Note that
the case $\kappa=0$ does not require modifications, since the
residue of the $R$-matrix vanishes at this point. At $\kappa=0$,
the bound state becomes weakly bound, and propagates far into the
bulk

The cases $\kappa>-1$, i.e. $h<h_c=2\sqrt{m}$, are also somewhat
special. In these cases, the R-matrix still has a pole on the
imaginary $\theta$ line, although not on the physical strip when
$\kappa\le0$. As noted in \cite{Ghoshal:1993tm}, the case
$\kappa=-1$ corresponds to a ``critical'' value of the magnetic
field, $h_c$, at which the reflection matrix happens to have a
third order zero at $\theta=0$. We will discuss the meaning of
this critical point from the viewpoint of the boundary
entanglement entropy in section \ref{discuss}.

Finally, we note also that the $R$ matrix at $\kappa=0$ is just
equal to the negative of the fixed-boundary condition $R$ matrix,
$\kappa=-\infty$, and that
\begin{eqnarray}
  R\left(\frac{i\pi}{2}-\theta\right) &=&
  -R\left(\frac{i\pi}{2}+\theta\right).\label{r1}
\end{eqnarray}

\section{Form factor expansion for the entanglement entropy in the boundary Ising model}
\subsection{Higher-particle form factors of Ising branch-point twist
fields}\indent \\

\noindent In order to evaluate the series (\ref{1p}) we need to
address first the issue of computing higher-particle form factors
of the twist fields in the Ising theory. In \cite{entropy} the
full set of form factor consistency equations was written but only
the two-particle form factors were explicitly calculated. However,
for the Ising theory, this will be the main piece of information
needed, as higher-particle form factors can be obtained out of
two-particle ones by using Wick's theorem.

Since the branch-point twist field is invariant under the internal
$\Z_2$ symmetry of the Ising model, characteristic of the Majorana
Fermions, only even-particle form factors will be non-zero. Let us
consider some of the consistency equations for the form factors of
the twist field $\mathcal{T}$ (with even number of particles $k$)
\begin{eqnarray}
  F_{k}^{\mathcal{T}|\ldots a_i a_{i+1} \ldots}(\ldots,\theta_i, \theta_{i+1}, \ldots ) &=&
  (-1)^{\delta_{a_ia_{i+1}}}F_{k}^{\mathcal{T}|\ldots a_{i+1} a_i\ldots }(\ldots,\theta_{i+1}, \theta_i,  \ldots ), \label{cpt}\\
 F_{k}^{\mathcal{T}|a_1 \ldots a_k}(\theta_1+2 \pi i, \ldots,
\theta_k) &=&
  F_{k}^{\mathcal{T}|a_2 \ldots a_k (a_1+1)}(\theta_2, \ldots, \theta_{k},
  \theta_1)
  \label{crossing}
\end{eqnarray}
where in we identify the particle types $a+n\equiv a$. Using these relations repeatedly,
it is possible to write all form factors in terms of form factors involving only one particle
type (say 1). For $1\leq a_j\leq n$ and with the ordering $a_1\ge \ldots \ge a_k$, we have
\beq\label{one}
    F_{k}^{\mathcal{T}|a_1 \ldots a_k}(\theta_1, \ldots,\theta_k)
    = F_{k}^{\mathcal{T}|1 \ldots 1}(\theta_1+2 \pi i(a_1-1), \ldots, \theta_k+2\pi i(a_k-1)),
\eeq and different orderings can be obtained using (\ref{cpt}), by
which extra signs may appear. Using this as a definition for form
factors with at least one particle of type different than 1, it is
possible to check that equations (\ref{cpt}) and (\ref{crossing})
are indeed satisfied, under the condition that (\ref{cpt}) holds
for all particles being of type 1: \beq\label{cptone}
  F_{k}^{\mathcal{T}|1\ldots 1}(\ldots,\theta_i, \theta_{i+1}, \ldots ) =
  -F_{k}^{\mathcal{T}|1\ldots 1 }(\ldots,\theta_{i+1}, \theta_i,  \ldots )
\eeq
and under one additional condition, coming from $n$ applications of (\ref{crossing}):
\beq\label{crossone}
    F_{k}^{\mathcal{T}|1 \ldots 1}(\theta_1+2 \pi i n , \ldots,\theta_k) =
    -F_{k}^{\mathcal{T}|1 \ldots 1}(\theta_1, \ldots,\theta_k)
\eeq
where we used the fact that the number of particles is even. That is, the set of equations (\ref{cpt}) and (\ref{crossing})
is consistent, and it is sufficient to solve (\ref{cptone}) and (\ref{crossone}).

Finally, there are two more conditions on form factors. We will write them in terms of
form factors with only particles of type 1. These are the kinematic residue equations:
\begin{eqnarray}
 -i \text{Res}_{\substack{\bar{\theta}_{0}={\theta}_{0}}}
 F_{k+2}^{\mathcal{T}|1 \ldots 1}(\bar{\theta}_0+i\pi,{\theta}_{0}, \theta_1 \ldots, \theta_k)
  =
  F_{k}^{\mathcal{T}|1 \ldots 1}(\theta_1, \ldots,\theta_k),\label{kre}
\end{eqnarray}
\begin{eqnarray}
 -i \text{Res}_{\substack{\bar{\theta}_{0}={\theta}_{0}}}
 F_{k+2}^{\mathcal{T}|1 1 1 \ldots 1}(\bar{\theta}_0+2i\pi n - i\pi,{\theta}_{0}, \theta_1 \ldots, \theta_k)
  =-
  F_{k}^{\mathcal{T}| 1 \ldots 1}(\theta_1, \ldots,\theta_k).\label{kre2}
\end{eqnarray}

Since we are dealing the free Fermion case, it is natural to
expect that the form factors of the twist field would admit closed
expressions in terms of Pfaffians, as for the order and disorder
fields of the Ising theory. This is indeed the case, and it is
easy to show that
\begin{equation}\label{f}
F_{k}^{\mathcal{T}|11\ldots 1}(\theta_1,
\ldots,\theta_{k})=\langle\mathcal{T}\rangle{\rm Pf}(\hat{K}),
\end{equation}
(recall that the Pfaffian has the property that ${\rm Pf}(\hat{K})^2 = \det(\hat{K})$)
where $\hat{K}$ is an anti-symmetric $k \times k$ matrix, $k$ even, with
entries
\begin{equation}\label{k}
    \hat{K}_{ij}=\frac{F_{\text{min}}^{\mathcal{T}|11}(\theta_i - \theta_j)}{F_{\text{min}}^{\mathcal{T}|11}(i \pi )P(\theta_i-\theta_j)}:=
    K(\theta_i-\theta_j),
\end{equation}
and
\begin{equation}
F_{\text{min}}^{\mathcal{T}|11}(\theta)=-i
\sinh\left(\frac{\theta}{2n}\right)\qquad \text{and}\qquad
P(\theta)=\frac{2n \sinh\left(\frac{i \pi +
\theta}{2n}\right)\sinh\left(\frac{i \pi -
\theta}{2n}\right)}{\sin\left(\frac{\pi}{n}\right)}.
\end{equation}
That is, these are the only non-vanishing form factors and are
general solutions to (\ref{cptone})-(\ref{kre2}). Note that the
Pfaffian expression is nothing else than the application of Wick's
theorem on the operators $Z_1(\theta)$ involved in the form
factors (\ref{ff}) (specialised to all particles being of type 1),
a contraction of $Z_1(\theta_1)$ with $Z_1(\theta_2)$ being
$K(\theta_1-\theta_2)$. Hereafter the following properties of the
function $K(\theta)$ will often be used:
\begin{eqnarray}
  K(\theta) &=& -K(-\theta), \label{p1}\\
  K(\theta)|_{n=1} &=& 0, \label{p2}\\
  \lt(K(\theta+ i s)\rt)^* &=& -K(\theta-is), \qquad \theta, s \in
  \mathbb{R},\label{p3}
\end{eqnarray}
where ``*" indicates complex conjugation.

It is easy to prove that (\ref{k}) solves
(\ref{cptone})-(\ref{kre2}). Equation (\ref{cptone}) simply means
that if we exchange two lines and the two corresponding (under the
transpose) columns of a matrix, its Pfaffian gets a minus sign.
Equation (\ref{crossone}) is a consequence of the property
$K(\theta+2i\pi n) = -K(\theta)$ and (\ref{p1}), and the fact that
if we change the sign of a line and the corresponding column of a
matrix, its Pfaffian also gets a minus sign. Finally, the
kinematic residue equation (\ref{kre}) can also be easily proved
from the structure of the Pfaffian, seeing it in terms of Wick's
theorem. Looking at the most singular term as
$\bar{\theta}_0\rightarrow \theta_0$, using the poles of
$K(\theta)$ itself at $\theta=\pm i\pi$ with residues $\pm i$, we
obtain (\ref{kre}).

\subsection{Two- and four-particle boundary corrections to the entanglement entropy} \label{twofour}

\subsubsection{Two-particle correction}

Employing (\ref{f}) for the two-particle form factor, it is not
difficult to evaluate the first correction to the saturation value
of the entanglement entropy (\ref{ent}) using (\ref{sell}) and
(\ref{2p}). We find
\begin{equation}\label{c1}
s_1(t,\kappa) = -\frac{1}{8}\int_{-\infty}^{\infty}  {d
\theta}\, \left( \frac{\kappa+\cosh\theta}
     {\kappa-\cosh\theta}\right)\left(\frac{\cosh\theta-1}{\cosh^2\theta}\right)e^{-t \cosh\theta},
\end{equation}
since
\begin{equation}\label{nf}
 -\frac{d}{d n} \left[ n K(-2\theta)\right]_{n=1}= \frac{\lt[F_{\text{min}}^{\mathcal{T}|11}(2\theta)\rt]_{n=1}
    }{\lt[F_{\text{min}}^{\mathcal{T}|11}(i \pi)\rt]_{n=1}}\frac{d}{d n}\left[ \frac{n}{P(2\theta)} \right]_{n=1}=
 \frac{i\pi}{2}\frac{\tanh\theta}{\cosh
\theta}.
\end{equation}
The correction $s_1(t,\kappa)$ is finite for all values of $t$,
including $t=0$, as can be seen in figure \ref{fig1}. At this
point it is possible to evaluate the integral above explicitly:
\begin{equation}\label{neg}
 c_1(\kappa) := s_1(0,\kappa)=\frac{1}{4} - \frac{\pi }{8} + \frac{\pi }{4 \kappa}
 - \frac{{\sqrt{1 -\kappa}}\,\left( \pi  + 2\arcsin(\kappa) \right) }
   {4\kappa{\sqrt{1 + \kappa}}},
\end{equation}
and in particular
\begin{equation}\label{m10}
 c_1(-1)=\frac{10-3\pi}{8} \qquad \text{and}\qquad c_1(0)=
\frac{\pi-2}{8}.
\end{equation}

In the case where $\kappa>0$, as was said in sub-section
\ref{sectRIsing}, the boundary state expression (\ref{bs}) needs
modifications due to the presence of a boundary bound state.
Following \cite{Ghoshal:1993tm}, the boundary state in the
multi-copy model has the expansion \beq
    |B\ket = \lt[1 + g_c \sum_{j=1}^n Z_j(0) + g_c^2\sum_{j,k=1}^n Z_j(0)Z_k(0) +
    \sum_{j=1}^n \int_{-\infty}^\infty \frac{d\theta}{4 \pi} \, R\lt(\frc{i\pi}2-\theta\rt) Z_j(-\theta) Z_j(\theta) + \cdots \rt]|0\ket
\eeq where $g_c$ is proportional to the residue of the $R$-matrix
at the bound-state pole. Since the branch-point twist fields have
zero one-particle form factors, the only possible modifications to
the result for $s_1(t,\kappa)$ above come from the quadratic term
$g_c^2\sum_{j,k=1}^n Z_j(0)Z_k(0)$. The correction to
$s_1(t,\kappa)$ would then be \beq  -g_c^2 e^{-t}\frac{d}{dn}
\left[\sum_{i,j=1}^n F_2^{\mathcal{T}|i
j}(0,0)\right]_{n=1}=-g_c^2 e^{-t}\frac{d}{dn}\left[n
\sum_{j=0}^{n-1} K(2\pi i j)\right]_{n=1}. \eeq Using the methods
of summation of \cite{entropy}, it is possible to show that this
correction vanishes. Hence $s_1(t,\kappa)$ is the correct leading
large-distance correction for all $\kappa\le 1$.

\subsubsection{Four-particle correction}
Let us consider now the four-particle boundary correction. From
(\ref{1p}) and (\ref{sell}) we find
\begin{equation}\label{s4}
    s_2(t,\kappa)=-\frac{1}{2}\left[\prod_{k=1}^{2}\int_{-\infty}^{\infty} \frac{d\theta_k}{4\pi}
    R\left(\frac{i\pi}{2}-\theta_{k}\right) e^{-t\cosh\theta_k}\right]\frac{d}{dn} \left[
    \sum_{i,j=1}^n \frc1{\bra{\cal T}\ket} F_{4}^{\mathcal{T}| ii jj}(-\theta_1, \theta_1, -\theta_2,
    \theta_2)
   \right]_{n=1},
\end{equation}
where
\begin{eqnarray}\label{sum}
     \sum_{i,j=1}^n F_{4}^{\mathcal{T}| iijj}(-\theta_1, \theta_1, -\theta_2,
    \theta_2)&=& n \sum_{j=1}^n F_{4}^{\mathcal{T}| 11 jj }(-\theta_1, \theta_1, -\theta_2,
    \theta_2) \nonumber \\
    &=& n \sum_{j=0}^{n-1}F_{4}^{\mathcal{T}| 11 11 }(-\theta_1, \theta_1, -\theta_2+ 2 \pi i j,
    \theta_2 + 2 \pi i j).
\end{eqnarray}
Here we have used (\ref{cpt}) and (\ref{one}). Employing
(\ref{f}) we find
\begin{eqnarray}
  && \frac{n}{\langle\mathcal{T}\rangle}\sum_{j=0}^{n-1}F_{4}^{\mathcal{T}| 11 11  }(-\theta_1, \theta_1, -\theta_2+ 2 \pi i j,
    \theta_2 + 2 \pi i j) \label{4}\\
    &&=n \sum_{a=0}^{n-1} \left(K(2 \theta_1) K(2 \theta_2)+K(\theta_{12}-2 \pi i a)K(\theta_{12}+2 \pi i a)+
    K(\hat{\theta}_{12}-2 \pi i a)K(-\hat{\theta}_{12}-2 \pi i a) \right) \nonumber \\
    &&=n^2 K(2 \theta_1) K(2 \theta_2)+ n\sum_{a=0}^{n-1} \left(K(\theta_{12}-2 \pi i a)K(\theta_{12}+2 \pi i
    a)-K(\hat{\theta}_{12}-2 \pi i a)K(\hat{\theta}_{12}+2 \pi i a)
    \right).\nonumber
\end{eqnarray}
Here and below we use $\theta_{ij} = \theta_i-\theta_j$ and
$\hat{\theta}_{ij}=\theta_i + \theta_j$. It is  simple to show
that the $n^2 K(2 \theta_1) K(2 \theta_2)$ term will give no
contribution to the derivative at $n=1$ (this is due to property
(\ref{p2})) so that only the terms in the sum remain. These terms
will give a contribution, since, employing (\ref{p3}), they can
actually be rewritten as
\begin{equation}\label{re}
    -n\sum_{a=0}^{n-1}
    \left[|K(\theta_{12}-2 \pi i a)|^2 -|K(\hat{\theta}_{12}-2 \pi i a)|^2
    \right].
\end{equation}
Writing things in this way is useful as we can employ one of our
main results in \cite{entropy}, namely that
\begin{equation}\label{main}
    \frac{d}{d n} \left[n \sum_{a=1}^{n-1}
    \left|K(\theta-2 \pi i a)\right|^2\right]_{n=1}=
    \frac{\pi^2}{2} \delta(\theta).
\end{equation}
Under integration in (\ref{s4}), the term with $\h{\theta}_{12}$ can be changed
into $\theta_{12}$, and the change of variable required
inverts the sign of $R(i\pi/2-\theta_2)$ using property (\ref{r1}).
Hence both terms give the same contribution. Thus we find that the four-particle correction to the saturation
value of the Ising entanglement entropy is given by
\begin{eqnarray}\label{s4t}
  s_2(t,\kappa) &=& \frac{1}{32}\int_{-\infty}^{\infty}\int_{-\infty}^{\infty} d\theta_1
  d\theta_2
  R\left(\frac{i\pi}{2}-\theta_1\right)R\left(\frac{i\pi}{2}-\theta_2\right)\delta(\theta_{12}) e^{-2t
  \cosh \frac{\theta_{12}}{2} \cosh \frac{\hat{\theta}_{12}}{2}},\\
   &=& \frac{1}{32}\int_{-\infty}^{\infty} d\theta
   R\left(\frac{i\pi}{2}-\theta\right)^2 e^{-2t \cosh \theta}=
   \frac{1}{32}\int_{-\infty}^{\infty} d\theta \left(\frac{\kappa+\cosh \theta}{\kappa - \cosh \theta}\right)^2
    \frac{1 - \cosh \theta}{1 + \cosh \theta}e^{-2t \cosh
   \theta}.\nonumber
\end{eqnarray}
Contrarily to the two-particle correction, we find that the
four-particle contribution is divergent as $t \rightarrow 0$. Technically,
the reason for this is that the integrand of (\ref{s4t}) is a function
that tends to the value $-1$ as $\theta \rightarrow \infty$ when
$t=0$. Therefore the integral at $t=0$ is divergent. In order to
find the precise behaviour of the correction as $t$ approaches 0,
one can rewrite the integral above as:
\begin{equation}\label{s4t2}
  s_2(t,\kappa) =
   \frac{1}{32}\int_{-\infty}^{\infty} d\theta \left[\left(\frac{\kappa+\cosh \theta}{\kappa - \cosh \theta}\right)^2
    \frac{1 - \cosh \theta}{1 + \cosh \theta}+1 \right]e^{-2t \cosh
   \theta} - \frac{1}{16} K_0 (2t).
\end{equation}
The behaviour of the Bessel function as $t$ goes to zero is
well-known,
\begin{equation}\label{bes}
    K_0(2t)= -\gamma-\log(t)+{O}(t^2\log t),
\end{equation}
where $\gamma=0.577216...$ is the Euler-Mascheroni constant. Written in this form, the integral part is now a finite constant
at $t=0$, and we may define
\begin{eqnarray}\label{fc}
 c_2(\kappa)&=&
    \frac{1}{16}\int_{0}^{\infty} d\theta \left[\left(\frac{\kappa+\cosh \theta}{\kappa - \cosh \theta}\right)^2
    \frac{1 - \cosh \theta}{1 + \cosh \theta}+1 \right] \nonumber\\
    &=&-\frac{1}{8(1+\kappa)^2} \left[ 2\kappa - 3\kappa^2 -1 +
  \frac{\kappa\left(2\kappa -1\right) (\pi+2\arcsin\kappa)}
   {{\sqrt{1 - \kappa^2}}}\right],
\end{eqnarray}
with in particular
\begin{equation}
    c_2(-1)=\frac{23}{120}.
\end{equation}
The constant $c_2(\kappa)$ is an increasing function of the magnetic field (i.e. a decreasing function of $\kappa$).
Therefore we have
\begin{equation}\label{uv4}
    s_2(t, \kappa) = \frac{1}{16}\log(t)+\frc{\gamma}{16} + c_2(\kappa)+{O}(t)
\end{equation}
(the order $O(t)$ comes from the next correction to the integral
part of (\ref{s4t2})). Note that in view of the short-distance
behaviour of the entanglement entropy (\ref{shla}), we expect that
the coefficients of the logarithmic divergencies at small $rm$
will add up to the finite number $1/12$ when all corrections are
considered. This will be proven in section \ref{exacsmall}. Also,
the constant $c_2(\kappa)$, like $c_1(\kappa)$ above, is a part of
the constant $V(\kappa)$ in (\ref{shla}); again, in principle one
should add up all such constants, for all corrections, in order to obtain $V(\kappa)$. This will be discussed also
in section \ref{exacsmall}.

An interesting mathematical phenomenon can be observed: although
the integral in (\ref{s4t2}) has the same value for
$\kappa=-\infty$ as for $\kappa=0$ for any $t>0$, we have
$c_2(-\infty) = 3/8$ different from $c_2(0) = 1/8$. The
explanation is that the limit $t\to0$ of the integral in
(\ref{s4t2}) as a function of $\kappa$ is not uniform. For all
values of $t>0$ we have $s_2(t,-\infty) = s_2(t,0)$, and there is
a maximum for $\kappa\in(-\infty,1)$ at a unique value
$\kappa=\kappa_0$. But as $t$ becomes smaller, the position of
this maximum shifts towards more negative values, until it reaches
$-\infty$ at $t=0$. There, if we take away the constant (as
function of $\kappa$) term $\frac{1}{16}\log(t)$ in order to make
the limit finite, the value of the maximum itself reaches
$c_2(-\infty)$. It is also possible to observe in the integral in
(\ref{fc}) that the symmetry between $\kappa=-\infty$ and
$\kappa=0$ is broken. Indeed, if $\kappa$ is very negative, the
term in parenthesis can be approximated by 1 except for values of
$\theta$ where $\kappa+\cosh\theta\approx0$. These are very large
values of $\theta$, but they are not damped by any other factor,
hence the mistake in approximating by 1 is non-negligible for any
$\kappa$.

This means that formally, the expansion (\ref{uv4}) is valid only for $\kappa>-\infty$.
For the case $\kappa=-\infty$, that is, the fixed boundary condition, we have to consider
the other order of the limits: first $\kappa\to-\infty$, then $t\to0$.
By the symmetry between $\kappa=-\infty$ and $\kappa=0$, we define
\beq
    c_2(-\infty) =: c_2(0) = \frc18
\eeq
so that (\ref{uv4}) still holds in the case of a fixed boundary condition, $\kappa=-\infty$.

In fact, it is instructive to obtain a more general small-$t$ expansion, where we take
simultaneously $\kappa\to-\infty$. Let us consider $t\to0$ with $-\kappa t = a$ fixed. We may use the change of variable
$s=\cosh\theta-1$ and write $s_2(t,\kappa)$ as
\beq
    \frc1{16} \int_0^{\infty} ds \lt(\frc{k+1+s}{k-1-s}\rt)^2 \lt( -\frc{\sqrt{s}}{(s+2)^{3/2}} + \frc1{s+1}\rt) e^{-2t(s+1)}
    - \frc1{16} \int_0^{\infty} ds \lt(\frc{k+1+s}{k-1-s}\rt)^2 \frc{e^{-2t(s+1)}}{s+1}.
\eeq
The first integral as a function of $\kappa$ has a uniform limit as $t\to0$ on $\kappa\in[-\infty,0)$, so that we can directly
take $\kappa=-\infty$ and $t=0$; this gives $(2-\log 2)/16$. The second integral does not have a uniform limit, but it can be evaluated
explicitly:
\[
    \frc1{16(a+t)}\lt( 4a e^{-2t} - (a+t)  \Gamma(0,2t) e^{2t} - 8a (a+t) e^{2(a+t)} \Gamma(0,2(a+t))\rt)
\]
where $\Gamma(z,u)$ is the incomplete Gamma function, $\int_u^\infty v^{z-1} e^{-v} dv$. The small-$t$ limit can then easily be taken:
\beq
    s_2(t,-a/t) = \frc1{16} \log(t) + \frac{\gamma}{16}+ c_2^\natural(a)+ O(t).
\eeq
where
\beq\label{c2nat}
    c_2^\natural(a) =  \frc{3}8 - \frc12 a e^{2a} \Gamma(0,2a) .
\eeq
It is easy to see that $c_2^\natural(a)$ interpolates between $\lim_{\kappa\to-\infty}c_2(\kappa)$
at $a=0$ to $c_2(-\infty)$ at $a=\infty$.

We expect this phenomenon of non-commutativity of the limits to be
generic: it should occur at all orders, except, as we have seen,
for the very first two-particle correction. Its meaning will be
explained in section \ref{discuss}. We also refer the reader to section \ref{discuss} for an analysis of the large-distance corrections found here.

\subsection{Higher particle boundary corrections to the entanglement entropy}\noindent\\

\noindent As one would expect, the form factor expressions which
are obtained from (\ref{f}) become more and more involved as the
number of particles is increased. In particular, it is easy to see
that the $2\ell$-particle form factor of $\mathcal{T}$ is made out
of the sum of
\begin{equation}\label{prod}
    1\times 3\times 5 \times \ldots \times (2\ell-1) =
    \frac{(2\ell-1)!}{2^{\ell-1}(\ell-1)!},
\end{equation}
terms. This number grows faster than exponentially with $\ell$.
However, as we have already seen for the 4-particle corrections,
not all of these terms contribute to the derivative at $n=1$ and
those that contribute give in many cases the same contribution due
to the fact that they are identical when integrated over. For
example, the two contributions in (\ref{re}) are actually
equivalent when integrated over $\theta_1$ and $\theta_2$. The
combination of these two factors, that is, terms whose derivative
vanishes at $n=1$ and terms that can be grouped together, allows
us to reduce very dramatically the amount of non-vanishing
contributions to the derivative that we obtain from higher
particle form factors.

The $2 \ell$-particle form factors contributing to the entanglement entropy (see (\ref{sell}) and (\ref{1p}))
occur in a sum of the form
\begin{eqnarray}
&& \sum_{j_1,\ldots,j_{\ell}=0}^{n}F_{2\ell}^{\mathcal{T}|j_1 j_1
\ldots j_\ell j_\ell }(-\theta_1, \theta_1, -\theta_2, \theta_2,
\ldots, -\theta_{\ell},\theta_{\ell})\nonumber\\
&&=
n\sum_{j_1,\ldots,j_{\ell-1}=0}^{n-1}F_{2\ell}^{\mathcal{T}|1\ldots
1}(-\theta_1, \theta_1, (-\theta_2)^{j_1}, \theta_2^{j_1}, \ldots,
(-\theta_{\ell})^{
j_{\ell-1}},\theta_{\ell}^{j_{\ell-1}}),\label{icila}
\end{eqnarray}
where
\begin{equation}\label{no}
\theta_a^b=\theta_a + 2 \pi i b.
\end{equation}
Here, we used (\ref{one}) and (\ref{cpt}), along with the fact that every copy number occurs in
pairs, so that no sign remains. From Wick's theorem, one of the terms contributing to this sum
will be the following contraction:
\begin{eqnarray}
F_{2\ell}^{\mathcal{T}|1\ldots 1}(-\theta_1, \theta_1,
(-\theta_2)^{j_1}, \theta_2^{j_1}, (-\theta_3)^{j_2},
\ldots,\theta_{\ell-2}^{ j_{\ell-3}},(-\theta_{\ell-1})^{
j_{\ell-2}},\theta_{\ell-1}^{ j_{\ell-2}}, (-\theta_{\ell})^{
j_{\ell-1}},\theta_{\ell}^{
j_{\ell-1}}),\tallcon{10}{2}{5}\tallcon{18}{2}{5}\tallcon{25}{2}{5}\tallcon{30}{2}{5}
\xtallcon{30}{1}{28}\label{this}
\end{eqnarray}
which corresponds to
\begin{eqnarray}
\sum_{j_1,\ldots,j_{\ell-1}=0}^{n-1}K((-\hat{\theta}_{12})^{j_1})
K(\hat{\theta}_{23}^{j_1-j_2})K(\hat{\theta}_{34}^{j_2-j_3})
\ldots K(\hat{\theta}_{\ell-1\,
\ell}^{j_{\ell-2}-j_{\ell-1}})K(\hat{\theta}_{\ell 1}^
 { j_{\ell-1}}).\label{only}
\end{eqnarray}
We will call such a term ``fully connected'', which denotes any contraction
where the sums over $j$'s cannot be factorised into a product of sums.

Thanks to (\ref{p2}), any fully connected term vanishes as $n\to1$. Below
we will show that the derivative with respect to $n$ is not zero as $n\to1$,
but converges to a distribution in the rapidities, generalising the main result
(\ref{main}) of \cite{entropy}. This means that the product of two or more fully connected
term has a vanishing derivative as $n\to1$, so that the only terms that will contribute
to (\ref{icila}) are the fully connected Wick contractions. We will further show
below that all fully connected terms can be brought to the form (\ref{this}).

\subsubsection{Explicit evaluation of fully connected terms}
The sum (\ref{only}) can be obtained in a systematic way by
exploiting a result which was first obtained in appendix 3 of
\cite{entropy}. The result derived there was a special case of the
sum
\begin{eqnarray}
\sum_{a=0}^{n-1}K((-x)^a)
  K(y^a) =-\frac{i\sinh\left(\frac{y+x}{2}\right)
}{2\cosh\frac{x}{2}
  \cosh\frac{y}{2}}\left(K(x+y-i\pi)+K(x+y + i\pi)\right),\label{sum2}
\end{eqnarray}
which can be computed in exactly the same manner. The sum
(\ref{only}) can be evaluated by simply using (\ref{sum2})
recursively. When doing so one realizes the need to distinguish
two special cases, depending on whether $\ell$ is even or odd in
(\ref{only}). The final expressions are,
\begin{eqnarray}
&&\sum_{j_1,\ldots,j_{2\ell-1}=0}^{n-1}K((-x_1)^{j_1})
  K(x_2^{j_1-j_2}) \ldots K(x_{2\ell-1}^{j_{2\ell-2}-j_{2\ell-1}})K(x_{2\ell}^{j_{2\ell-1}})=
   \frac{(-1)^{\ell} 2 i \sinh(\frac{\sum_{i=1}^{2\ell}x_i}{2})}{\prod\limits_{i=1}^{2 \ell} 2
  \cosh\frac{x_i}{2}}\nonumber \\
  && \qquad \quad\times \sum_{j=1}^{\ell} \left(%
\begin{array}{c}
  2\ell-1 \\
  \ell-j \\
\end{array}%
\right)
  \left[K(\sum_{i=1}^{2\ell}x_i + (2j-1) \pi i)+K(\sum_{i=1}^{2\ell}x_i - (2j-1) \pi
  i)\right],\label{for}
\end{eqnarray}
and
\begin{eqnarray}
&&\sum_{j_1,\ldots,j_{2\ell}=0}^{n-1}K((-x_1)^{j_1})
  K(x_2^{j_1-j_2})\ldots K(x_{2\ell}^{j_{2\ell-1}-j_{2\ell}})K(x_{2\ell+1}^{j_{2\ell}})=
   \frac{(-1)^{\ell+1} 2 \cosh(\frac{\sum_{i=1}^{2\ell+1}x_i}{2})}{\prod\limits_{i=1}^{2 \ell+1} 2
  \cosh\frac{x_i}{2}}\nonumber \\
  &&\times \left[\left(%
\begin{array}{c}
  2\ell \\
  \ell \\
\end{array}%
\right)K(\sum_{i=1}^{2\ell+1}x_i)+\sum_{j=1}^{\ell} \left(%
\begin{array}{c}
  2\ell \\
  \ell-j \\
\end{array}%
\right)
  \left[K(\sum_{i=1}^{2\ell+1}x_i + 2\pi i j)+K(\sum_{i=1}^{2\ell+1}x_i - 2 \pi
  i j)\right] \right].\label{for2}
\end{eqnarray}
A proof by induction of these identities is provided in appendix
\ref{proof}.

\subsubsection{Analytic continuation in $n$ and computation of the derivative at $n=1$ of fully connected terms}

Let us start by considering the analytic continuation and
evaluating the derivative at $n=1$ of (\ref{for}). This summation
formula already provides an analytic continuation in $n$ of the
multiple sum over $j$'s for any fix $x$'s. Naturally, such an
analytic continuation is not unique. Additionally, this formula
itself gives many more analytic continuations when taken under
integration over the $x$ variables. Indeed, for integer $n$, all
poles of the functions $K$ involved are exactly cancelled by the
zeros of $\sinh(\frac{\sum_{i=1}^{2\ell}x_i}{2})$, hence the
integration contours can be moved away from the real $x$-axis
without any change to the answer. But for non-integer $n$, poles
and zeros generically are at different points, except for the
poles of $K$ at values of its argument $\pm i\pi$. Hence,
different integration contours give different analytic
continuations of the integrals.

Below we suggest that the formula (\ref{for}) is the correct one, but that
the contours need to be taken differently. Let us consider the rapidity variables $\theta_i$, with $x_i = \h\theta_{i,i+1}$
(and $x_{2\ell} = \theta_{2\ell,1}$).
For $n$ large enough, all the poles of $K$ are beyond the integration contour of the variable
\beq
    \theta = \sum_i \theta_i = \frc12\sum_i x_i
\eeq
for all $j$'s in (\ref{for}). However, as $n$ decreases, poles cross the $\theta$ integration contour. At the points where
these poles cross, there is a zero provided by the hyperbolic sine function, so that no discontinuity occurs. Yet as function of
$n$, the result is not smooth, since there is a discontinuity in the {\em derivative} with respect to $n$. This is because
as $n$ approaches an integer value where a pole crosses the integration contour, the right-hand side of
(\ref{for}) is not uniformly convergent as function of $\theta$, and develops an infinitely thin peak of finite height
at $\theta=0$. There is no natural way of modifying (\ref{for}) in order to avoid this phenomenon. Indeed, it is
related to the fact that for integer values of $n$, the unambiguous value of the sum in (\ref{for}) at $\theta=0$ varies
with $n$ up to $n=\ell+1$, but from $n=\ell+1$ up to infinity it takes the same values. Hence, any analytic continuation
will have to reproduce this unnatural behaviour. We note that this phenomenon is a generalisation of that observed
in \cite{entropy} in the two-particle approximation of the bulk two-point function (and above in the four-particle boundary case):
there the value of the sum at $\theta=0$ at $n=2,3,4,\ldots$ is constant and non-zero,
while the value at $n=1$ is zero. The resulting non-uniform convergence was at the basis of the calculation of the
derivative with respect to $n$ at $n=1$.

In order to recover a smooth function of $n$ up to $n=1$, we then
need to move the $\theta$ integration contour towards values where
the argument of $K$ is just $\theta\pm i\pi$ in (\ref{for}), for
all $j$'s. It is convenient to still avoid poles of the hyperbolic
cosine factors in the denominator. Hence, we consider shifting all
$\theta_i$ by $\mp\frc{j-1}{2\ell}\pi i$, so that we get the
following summation formula, valid under integration:
\begin{eqnarray}
&&\sum_{j_1,\ldots,j_{2\ell-1}=0}^{n-1}K((-\hat{\theta}_{12})^{j_1})
  K(\hat{\theta}_{23}^{j_1-j_2})\ldots K(\hat{\theta}_{
  2\ell, 1}^{j_{2\ell-1}})\stackrel{\int}=
    2 i \sinh\theta
\nonumber \\
  && \qquad \quad\times \sum_{j=1}^{\ell} \sum_{q=\pm} (-1)^{\ell+j-1} \left(%
\begin{array}{c}
  2\ell-1 \\
  \ell-j \\
\end{array}%
\right)
  \frc{K(2\theta+ q \pi i) \prod\limits_{i=1}^{2\ell}\shift_{\theta_i\to\theta_i- q \frc{j-1}{2\ell}\pi i}}{\prod\limits_{i=1}^{2 \ell} 2
  \cosh\lt(\frac{\h\theta_{i,i+1}}{2} - q \frc{j-1}{2\ell} \pi i\rt)}.\label{forgood}
\end{eqnarray}
Here, $\shift_{\theta_i\to\theta_i-\frc{j-1}{2\ell}\pi i}$ is an operator acting on all other functions in the integrand (that is, those not appearing
here), indicating that their arguments must be modified as written. More precisely, this indicates a shift of contour, and if the other functions
in the integrand have poles that are crossed by this shift, then the residues must be taken. It is important, of course, that the integrals
over rapidities stay convergent at all stages of the shifts.

The strongest evidence for the validity of this formula is provided by the fact that in the bulk case, the
scaling dimension of the twist fields can be reproduced by re-summing the form factor expansion,
for any real $n>1$. In appendix \ref{scaldim} we provide very convincing numerics for this scaling dimension.

We can now evaluate the derivative with respect to $n$ at $n=1$ of this. For generic $\theta$, the function
on the right-hand side of (\ref{forgood}) is zero at $n=1$, but at $\theta=0$, it is non-zero. Hence, we need to properly
take the limit $n\to1$ of the derivative of the right-hand side of (\ref{forgood}) as a distribution. We have
\beq
    -\frc{d}{dn} \lt[\sinh\theta\,K(2\theta\pm \pi i)\rt]_{n=1} = \mp \frc{\pi}2 \frc{\cosh\theta}{\sinh\theta},
\eeq
but this has a pole at $\theta=0$, coming from the kinematic pole of $K$ at $\pm (2n-1) \pi i$. It can be resolved
by noticing that for $\theta\to0$ and $n\to1$, we have
\beq
    -\frc{d}{dn} \lt[\sinh\theta\,K(2\theta\pm \pi i)\rt] \sim \mp \frc{\pi}{2} \frc1{\theta\mp (n-1) \pi i}.
\eeq
Hence, as a distribution,
\beq
    -\frc{d}{dn} \lt[\sinh\theta\,K(2\theta\pm \pi i)\rt]_{n=1} = \mp \frc{\pi}2 \frc{\cosh\theta}{\sinh(\theta\mp i 0^+)}
\eeq
and this gives
\begin{eqnarray}
&&-\frac{d}{dn}\left[\sum_{j_1,\ldots,j_{2\ell-1}=0}^{n-1}K((-\hat{\theta}_{12})^{j_1})
  K(\hat{\theta}_{23}^{j_1-j_2})\ldots K(\hat{\theta}_{
  2\ell, 1}^{j_{2\ell-1}})\right]_{n=1}\stackrel{\int}=
    \pi i \cosh\theta
\nonumber \\
  && \qquad \quad\times \sum_{j=1}^{\ell} \sum_{q=\pm} (-1)^{\ell+j} q \left(%
\begin{array}{c}
  2\ell-1 \\
  \ell-j \\
\end{array}%
\right)
  \frc{{\rm csch}(\theta-q i 0^+) \prod\limits_{i=1}^{2\ell}\shift_{\theta_i\to\theta_i-q\frc{j-1}{2\ell}\pi i}}{\prod\limits_{i=1}^{2 \ell} 2
  \cosh\lt(\frac{\h\theta_{i,i+1}}{2} -q\frc{j-1}{2\ell} \pi i\rt)}.
\end{eqnarray}
Finally, we may simplify this formula by shifting back the contours towards their initial positions. Doing so,
the integrand will be zero, and the only contributions will be poles taken on
the way. The final result is
\begin{eqnarray}
&&-\frac{d}{dn}\left[\sum_{j_1,\ldots,j_{2\ell-1}=0}^{n-1}K((-\hat{\theta}_{12})^{j_1})
  K(\hat{\theta}_{23}^{j_1-j_2})\ldots K(\hat{\theta}_{
  2\ell, 1}^{j_{2\ell-1}})\right]_{n=1}\stackrel{\int}=
    \\
  && -2\pi^2 \sum_{j=1}^{\ell}\sum_{k=1}^{j} \sum_{q=\pm} (-1)^{\ell+j}
\lt\{\ba{cc} 1/2 & (k=j) \z 1 & (k<j) \ea\rt\}  \left(%
\begin{array}{c}
  2\ell-1 \\
  \ell-j \\
\end{array}%
\right)
    \frc{\prod\limits_{i=1}^{2\ell}\shift_{\theta_i\to\theta_i+ q\frc{j-k}{2\ell}\pi i}}{
    \prod\limits_{i=1}^{2 \ell} 2
  \cosh\lt(\frac{\h\theta_{i,i+1}}{2} + q\frc{j-k}{2\ell} \pi i\rt)} \delta(\theta). \nonumber
\end{eqnarray}
The case $k=j$ in this formula can be simplified using
\beq
  -\frac{ 2 \pi^2 \delta(\theta)\sum\limits_{j=1}^{\ell} \left(%
\begin{array}{c}
 2\ell-1 \\
  \ell-j \\
\end{array}%
\right)
  (-1)^{j+\ell}}{\prod\limits_{j=1}^{2\ell} 2
  \cosh\frac{\hat{\theta}_{j j+1}}{2}}
  = \frac{ \left(%
\begin{array}{c}
 2\ell-2 \\
  \ell-1 \\
\end{array}%
\right)(-1)^{\ell} 2 \pi^2 \delta(\theta)}{
 \prod\limits_{j=1}^{2\ell} 2
  \cosh\frac{\hat{\theta}_{j j+1}}{2}},
\eeq
so that we obtain
\begin{eqnarray} \label{term}
&&-\frac{d}{dn}\left[\sum_{j_1,\ldots,j_{2\ell-1}=0}^{n-1}K((-\hat{\theta}_{12})^{j_1})
  K(\hat{\theta}_{23}^{j_1-j_2})\ldots K(\hat{\theta}_{
  2\ell, 1}^{j_{2\ell-1}})\right]_{n=1}\stackrel{\int}=
    \\
  && (-1)^\ell 2\pi^2  \delta(\theta)\lt[
\frac{ \left(%
\begin{array}{c}
 2\ell-2 \\
  \ell-1 \\
\end{array}%
\right) }{
 \prod\limits_{j=1}^{2\ell} 2
  \cosh\frac{\hat{\theta}_{j j+1}}{2}} -
 \sum_{j=1}^{\ell}\sum_{k=1}^{j-1} \sum_{q=\pm}
    \frc{
 \left(%
\begin{array}{c}
  2\ell-1 \\
  \ell-j \\
\end{array}%
\right) (-1)^{j}\prod\limits_{i=1}^{2\ell}\shift_{\theta_i\to\theta_i+ q\frc{j-k}{2\ell}\pi i}}{
    \prod\limits_{i=1}^{2 \ell} 2
  \cosh\lt(\frac{\h\theta_{i,i+1}}{2} + q\frc{j-k}{2\ell} \pi i\rt)} \rt]. \nonumber
\end{eqnarray}
Note that the first term inside the square brackets on the
right-hand side is what is obtained by directly using (\ref{for}),
without contour shifts; it is a direct generalisation of the
four-particle case. The other terms are corrections,
characteristic of higher-particle contributions only.

A similar analysis can be made for the odd case, (\ref{for2}). The result is
\begin{eqnarray} \label{term2}
&&-\frac{d}{dn}\left[\sum_{j_1,\ldots,j_{2\ell}=0}^{n-1}K((-\hat{\theta}_{12})^{j_1})
  K(\hat{\theta}_{23}^{j_1-j_2})\ldots K(\hat{\theta}_{
  2\ell+1, 1}^{j_{2\ell}})\right]_{n=1}\stackrel{\int}=
    \\
  &&\qquad\qquad  (-1)^{\ell} 2\pi^2  \delta(\theta)
 \sum_{j=1}^{\ell}\sum_{k=1}^{j} \sum_{q=\pm}
    \frc{
 \left(%
\begin{array}{c}
  2\ell \\
  \ell-j \\
\end{array}%
\right) (-1)^{j}q \prod\limits_{i=1}^{2\ell+1}\shift_{\theta_i\to\theta_i+ q\frc{j-k+1/2}{2\ell+1}\pi i}}{
    \prod\limits_{i=1}^{2 \ell+1} 2
  \cosh\lt(\frac{\h\theta_{i,i+1}}{2} + q\frc{j-k+1/2}{2\ell+1} \pi i\rt)}. \nonumber
\end{eqnarray}
Note that in this case, we get ``pure correction terms'', as directly taking the derivative with respect to $n$ at $n=1$
of (\ref{for2}), without contour shifts, gives zero.

The final results as written in (\ref{term}) and (\ref{term2})
hold only if the other functions of the integrand do not have
poles on the region covered by the shifts; then the shift
operators are just shift of arguments. The contributions of poles
may be evaluated in a similar way, but below we will avoid these
complications. Also, these formulae are valid for
$\ell=1,2,3,\ldots$ only, that is, excluding $\ell=0$.

\subsubsection{Putting everything together}

The final step in order to evaluate $s_{\ell}(t,\kappa)$ is to
work out the ``multiplicity" of (\ref{only}) for a fixed particle
number. This can be easily done by exploiting the Wick contraction
picture introduced in (\ref{this}). Let us pick the first rapidity
$-\theta_1$. There are $2(\ell-1)$ possible contractions that
could be performed as a contraction with $\theta_1$ is not allowed
(that would produce a non-fully connected term). Let us assume that $-\theta_1$ is
connected to $\theta_i^j$ for some fixed $i, j$. Then, if we now
pick the next rapidity, that is $\theta_1$, it can be connected to
almost any term, except for $-\theta_i^j$  (that would again be
non-fully connected) and the two that are already connected. That
gives us $2(\ell-2)$ possibilities. We carry on this argument
by connecting $\theta_i^j$ to some $\theta_{-i'}^{j'}$, then
$\theta_{i'}^{j'}$ to some $\theta_{-i''}^{j''}$, etc., until no rapidities
are left. Hence we find that the total number of
fully connected terms is $(\ell-1)! 2^{\ell-1}$. All these terms
are identical when integrated in all rapidities because they can all be brought
to the form (\ref{only}) by a series of two types of operation. First, we may change the sign
of a rapidity, getting a minus sign from the $R$ matrix thanks to property (\ref{r1}),
then change the order of the two $Z$ operators
associated to this rapidity, which cancels this minus sign. Second, we may move pairs of $Z$ operators
associated to a given rapidity without getting any sign.

In order to be able to use formulae (\ref{term}) and (\ref{term2}), we must make sure that no poles
occur in the regions covered by the shifts, for any $\ell$.
This imposes that the other functions in the integrand should not have poles for ${\rm Im}(\theta_i) \in (-\pi/2,\pi/2)$.
Since factors $R(i\pi/2-\theta_i)$ occur in the integrand, we must take
\beq
    \kappa\leq 0.
\eeq
With this restriction, we obtain
\begin{eqnarray}
    s_{2\ell}(t,\kappa) &=&
    \frac{(2\ell-1)! 2^{2\ell-1}}{(2\ell)!
    (4\pi)^{2\ell}}\left[\prod_{k=1}^{2\ell} \int_{-\infty}^{\infty} d \theta_{k}e^{-t \cosh \theta_{k}}
    R\left(\frac{i\pi}{2}-\theta_{k}\right)\right] \n && \qquad \times
    \lt(-\frac{d}{dn}\rt)\left[\sum_{j_1,\ldots,j_{2\ell-1}=0}^{n-1}K(-\hat{\theta}_{12}^{j_1})
  K(\hat{\theta}_{23}^{j_1-j_2})\ldots K(\hat{\theta}_{
  2\ell, 1}^{j_{2\ell-1}})\right]_{n=1} \n
    &=&  \label{1ps}
    \frac{ \pi^2 (-1)^{\ell} }{2
    \ell}\left[\prod_{k=1}^{2\ell} \int_{-\infty}^{\infty} \frac{d \theta_{k}}{4\pi}\rt]
     \delta(\theta)\lt[
\left(%
\begin{array}{c}
 2\ell-2 \\
  \ell-1 \\
\end{array}%
\right) \prod\limits_{j=1}^{2\ell}  \frc{e^{-t \cosh \theta_{j}} R\left(\frac{i\pi}{2}-\theta_{j}\right)}{
    \cosh\frac{\hat{\theta}_{j j+1}}{2}} \rt. \\ &&  \lt. -
 \sum_{j=1}^{\ell}\sum_{k=1}^{j-1} \sum_{q=\pm}
 \left(%
\begin{array}{c}
  2\ell-1 \\
  \ell-j \\
\end{array}%
\right)     (-1)^{j} \prod\limits_{i=1}^{2\ell}  \frc{e^{-t \cosh (\theta_{i}+ q\frc{j-k}{2\ell}\pi i)}
 R\left(\frac{i\pi}{2}-(\theta_{i}+ q\frc{j-k}{2\ell}\pi i)\right)}{
  \cosh\lt(\frac{\h\theta_{i,i+1}}{2} + q\frc{j-k}{2\ell} \pi i\rt)} \rt] \no
\end{eqnarray}
and similarly
\begin{eqnarray}\label{1ps2}
    && s_{2\ell+1}(t,\kappa)  = \frac{ \pi^2 (-1)^{\ell} }{2
    \ell+1}\left[\prod_{k=1}^{2\ell+1} \int_{-\infty}^{\infty} \frac{d \theta_{k}}{4\pi}\rt]\delta(\theta) \\ &&\qquad \times
  \sum_{j=1}^{\ell}\sum_{k=1}^{j} \sum_{q=\pm}
 \left(%
\begin{array}{c}
  2\ell \\
  \ell-j \\
\end{array}%
\right)     (-1)^{j} q \prod\limits_{i=1}^{2\ell+1}  \frc{e^{-t \cosh (\theta_{i}+ q\frc{j-k+1/2}{2\ell+1}\pi i)}
 R\left(\frac{i\pi}{2}-(\theta_{i}+ q\frc{j-k+1/2}{2\ell+1}\pi i)\right)}{
  \cosh\lt(\frac{\h\theta_{i,i+1}}{2} + q\frc{j-k+1/2}{2\ell+1} \pi i\rt)}, \no
\end{eqnarray}
both formulae hold for $\ell=1,2,3,\ldots$. All integrals involved
are absolutely convergent, and it is easy to see that this holds
throughout all shifts required. Notice that for $\ell=1$ in
(\ref{1ps}) we recover the result (\ref{s4t}) as it should be.
Setting $\ell=0$ in (\ref{1ps2}) does not give (\ref{c1}), since
$\ell=0$ is out of the range of applicability of these formulae;
the two-particle case $s_1(t,\kappa)$ is a special case.

Hence, we have for the entanglement entropy in the boundary case
\beq\label{boundaryfinal}
    S_A^{\rm boundary}(rm) = -\frc1{12} \log(\varep m) + \frc{U}2 + \sum_{\ell=1}^\infty s_\ell(2rm,\kappa)
\eeq
with (\ref{c1}), (\ref{1ps}) and (\ref{1ps2}).

Formulae (\ref{1ps}) and (\ref{1ps2}) are quite lengthy, but can be written in a more symmetric way,
more appropriate for numerical calculations. We present such alternative expressions in appendix \ref{Formulae}.

\section{Form factor expansion for the bulk entanglement entropy}
\label{sectbulk}

We have studied the entropy of the Ising model in the presence of
a boundary. However the form factor approach employed so far has
first been used for bulk theories in \cite{entropy} where the
entropy was obtained in the two-particle approximation. We should
mention here that the entanglement entropy of free theories was
obtained by a different method in \cite{casini1, casini2}. There
it was shown how it is connected to Painlev\'e transcendents. The
analysis of the previous sections can be easily adapted now to
find closed expressions for all entropy contributions in the bulk
case. Employing a form factor expansion for the two-point
functions of the twist fields in the bulk theory we can write
\begin{eqnarray}
&&\langle \mathcal{T}(r) \tilde{\mathcal{T}}(0)\rangle
-\langle\mathcal{T}\rangle^2 =
\sum_{\ell=1}^{\infty}\frac{1}{(2\ell)!}\sum_{j_1,\ldots,j_{2\ell}=1}^n\left[
\prod_{j=1}^{2\ell}\int\limits_{-\infty }^{\infty } \frac{d\theta
_{j}}{ 2\pi}\,e^{-rm
\cosh\theta_{j}}\right]\left|F_{2\ell}^{\mathcal{T} |j_1 j_2
\ldots
j_{2\ell}}(\theta _{1},\ldots,\theta_{2\ell})\right|^2\nonumber\\
&&=\sum_{\ell=1}^{\infty}\frac{n}{(2\ell)!}\sum_{j_1,\ldots,j_{2\ell-1}=0}^{n-1}\left[
\prod_{j=1}^{2\ell}\int\limits_{-\infty }^{\infty } \frac{d\theta
_{j}}{2\pi}\,e^{-rm
\cosh\theta_j}\right]\left|F_{2\ell}^{\mathcal{T} |1 \ldots
1}(\theta
_{1},\theta_2^{j_1},\ldots,\theta_{2\ell}^{j_{2\ell-1}})\right|^2.\label{ent11}
\end{eqnarray}
Since we want to compute the derivative at $n=1$ of the above, we
need to evaluate
\begin{eqnarray}\label{dn2}
   - \frac{d}{dn}\left[n \sum_{j_1,\ldots,j_{2\ell-1}=0}^{n-1} \left|F_{2\ell}^{\mathcal{T} |1 \ldots
1}(\theta
_{1},\theta_2^{j_1},\ldots,\theta_{2\ell}^{j_{2\ell-1}})\right|^2\right]_{n=1}.
\end{eqnarray}
Notice that, for the free Fermion theory
\begin{eqnarray}
\left|F_{2\ell}^{\mathcal{T} |1 \ldots 1}(\theta
_{1},\theta_2^{j_1},\ldots,\theta_{2\ell}^{j_{2\ell-1}})\right|^2
=(-1)^{\ell}F_{2\ell}^{\mathcal{T} |1 \ldots 1}(\theta
_{1},\theta_2^{j_1},\ldots,\theta_{2\ell}^{j_{2\ell-1}})F_{2\ell}^{\mathcal{T}
|1 \ldots 1}(\theta
_{1},\theta_2^{-j_1},\ldots,\theta_{2\ell}^{-j_{2\ell-1}}),
\end{eqnarray}
where we used (\ref{p3}).

The only contribution to the derivative
(\ref{dn2}) will again come from the fully connected terms for
similar reasons as in the boundary case. In fact, it is not
difficult to convince oneself that the non-vanishing contributions
coming from the $2\ell$-particle form factor will be completely
analogous to the non-vanishing contributions coming from the
$4\ell$-particle form factor in the boundary case, the only
difference being the presence of the reflection matrices in the
latter case and the amount of terms that contribute.
Let us consider a very particular fully connected
term, corresponding to the Wick contractions
\begin{equation}
F_{2\ell}^{\mathcal{T} |1 \ldots 1}(\theta
_{1},\theta_2^{j_1},\theta_3^{j_2},\theta_4^{j_3},\ldots,\theta_{2\ell-1}^{j_{2\ell-2}},\theta_{2\ell}^{j_{2\ell-1}})
\tallcon{11}{2}{3}\tallcon{14}{2}{3}\tallcon{6}{2}{4}F_{2\ell}^{\mathcal{T}
|1 \ldots 1}(\theta
_{1},\theta_2^{-j_1},\theta_3^{-j_2},\ldots,\theta_{2\ell-2}^{-j_{2\ell-3}},
\theta_{2\ell-1}^{-j_{2\ell-2}},\theta_{2\ell}^{-j_{2\ell-1}})\tallcon{10}{2}{5}\tallcon{17}{2}{4}
\xtallcon{17}{1}{15}. \label{prod2}
\end{equation}
The contribution of this term to (\ref{dn2}) is
\begin{eqnarray}\label{dn3}
&& - \frac{d}{dn}\left[n (-1)^\ell
\sum_{j_1,\ldots,j_{2\ell-1}}
\left(K(\theta_{12}^{-j_1})\prod_{k=1}^{\ell-1}K(\theta_{2k+1,
2k+2}^{j_{2k}-j_{2k+1}})\right) \left(K(\theta_{1,2\ell}^{
j_{2\ell-1}} )\prod_{k=1}^{\ell-1}K(\theta_{2k,
2k+1}^{-j_{2k-1}+j_{2k}})\right) \right]_{n=1}\nonumber \\
&& =- \frac{d}{dn}\left[n \sum_{j_1,\ldots,j_{2\ell-1}}
\left(K((-\theta_{12})^{j_1})K(\theta_{1,2\ell}^{ j_{2\ell-1}}
)\prod_{k=1}^{\ell-1}K(\theta_{2k+1,
2k+2}^{j_{2k}-j_{2k+1}})K((-\theta_{2k,
2k+1})^{j_{2k-1}-j_{2k}})\right) \right]_{n=1}. \nonumber \eeqa By
changing the sign of rapidities with odd index, which does not
change the result under integration in (\ref{ent11}), we obtain
exactly (\ref{term}).

As for the boundary case, it is possible to argue that all fully
connected terms are identical to the one above when integrated in
all rapidities. The number of such terms can easily be evaluated
by the following argument: Let us consider the first form factor
in the product (\ref{prod}) and pick one of the rapidity variables
on which it depends, say $\theta_i^j$ for some fixed $i, j$. It
can be connected to any other rapidity in the same form factor,
say $\theta_{i'}^{j'}$, and
therefore there are $2\ell-1$ possibilities. Now we can look at the
second form factor, picking $\theta_{i'}^{-j'}$. There are
$2\ell-2$ possible connections, since it cannot be connected to
$\theta_i^{-j}$ as this would produce a factorisable term. Suppose it is
connected to $\theta_{i''}^{-j''}$. We now come back to the first form factor,
looking at $\theta_{i''}^{j''}$. It can be connected to any other available
rapidities, so there are $2\ell-3$. Continuing, we find that there are
$(2\ell-1)!$ fully connected terms. Under the integral all these terms
are equivalent, because exchanging two rapidities in both factors simultaneously
does not bring out any sign. Therefore,
\beq
    -\frac{d}{dn}\left[\langle \mathcal{T}(r)
    \tilde{\mathcal{T}}(0)\rangle -\langle\mathcal{T}\rangle^2
    \right]= \sum_{\ell=1}^{\infty} e_\ell(rm)
\eeq
with
\begin{eqnarray}\label{el}
    && e_{\ell}(rm) =
    \frac{ \pi^2 (-1)^{\ell} }{
    \ell}\left[\prod_{k=1}^{2\ell} \int_{-\infty}^{\infty} \frac{d \theta_{k}}{4\pi}\rt]\delta(\theta) \\ &&  \times
     \lt[
\left(%
\begin{array}{c}
 2\ell-2 \\
  \ell-1 \\
\end{array}%
\right) \prod\limits_{j=1}^{2\ell}  \frc{e^{-rm \cosh \theta_{j}} }{
    \cosh\frac{\hat{\theta}_{j j+1}}{2}} -
 \sum_{j=1}^{\ell}\sum_{k=1}^{j-1} \sum_{q=\pm}
 \left(%
\begin{array}{c}
  2\ell-1 \\
  \ell-j \\
\end{array}%
\right)     (-1)^{j} \prod\limits_{i=1}^{2\ell}  \frc{e^{-rm \cosh (\theta_{i}+ q\frc{j-k}{2\ell}\pi i)}
 }{\cosh\lt(\frac{\h\theta_{i,i+1}}{2} + q\frc{j-k}{2\ell} \pi i\rt)} \rt] \no
\end{eqnarray}
so that
\beq\label{bulkfinal}
    S_A^{\rm bulk}(rm) = -\frc16 \log(\varep m) + U + \sum_{\ell=1}^\infty e_\ell(rm).
\eeq
Again, see appendix \ref{Formulae} for an alternative expression of $e_\ell(rm)$.
In particular, the $\ell=1$ contribution is given by
\begin{equation}
e_1(rm)=-\int\limits_{-\infty }^{\infty } \int\limits_{-\infty
}^{\infty }\frac{d\theta_1 d\theta_2}{16}\,e^{-rm(
\cosh\theta_{1}+\cosh\theta_2)} \frac{ \delta(\h\theta_{12})}{
  \cosh^2\frac{{\h\theta}_{12}}{2}}= -\int\limits_{-\infty }^{\infty
}\frac{d\theta}{16}\,e^{-2rm\cosh\theta}=-\frac{K_0(2rm)}{8},
  \end{equation}
which is one of the main results obtained in \cite{entropy}.

\section{Exact UV behaviour of the entanglement entropy in the boundary Ising model}
\label{exacsmall}

\subsection{Exact logarithmic behaviour}\label{exactlogbound}\indent \\

\noindent Let us start by extracting the exact small-$rm$
logarithmic behaviour of the bulk entanglement entropy from the
full form factor expansion (\ref{bulkfinal}). For this purpose,
after integrating $\theta_{2\ell}$ using the delta-function, it is
convenient to change variables to the set $x_i=\h\theta_{i,i+1}$
for $i=1,\ldots,2\ell-2$, and $\theta_{2\ell-1}$. Then, we have
\beq\label{lecas}
    \theta_i = \sum_{j=i}^{2\ell-2} (-1)^{j-i} x_j + (-1)^{1+i} \theta_{2\ell-1}
\eeq
and in particular,
\beq
    \sum_{i=1}^{2\ell-1}\theta_i = \sum_{j=1}^{\ell-1} x_{2j-1} +\theta_{2\ell-1}, \quad
    \sum_{i=1}^{2\ell-2} \theta_i = \sum_{j=1}^{\ell-1} x_{2j-1} , \quad
    \sum_{i=2}^{2\ell-1} \theta_i = \sum_{j=2}^{\ell-1} x_{2j}. \label{lorignal}
\eeq
Hence, the hyperbolic cosine factors in the denominators, involving all $x_i$'s as well as the last two
sums above, do not depend on $\theta_{2\ell-1}$, so that it is
the large $|\theta_{2\ell-1}|$ behaviour of the integrand that determines the singularity at small $rm$.
In the exponential, we have at large $|\theta_{2\ell-1}|$ and fix $x_i$'s,
\beq
    \sum_{i=1}^{2\ell-1} \cosh\theta_i + \cosh \lt(\sum_{i=1}^{2\ell-1} \theta_i\rt)
    \propto e^{|\theta_{2\ell-1}|}.
\eeq Hence, the exponential factor will display a jump of finite
width from 1 to 0 around $|\theta_{2\ell-1}| \sim -\log(rm) +
const.$ when $rm$ is small, so that the leading small-$rm$
behaviour can be obtained by omitting the exponential factor and
replacing $\int d\theta_{2\ell-1}$ by $-2\log(rm)$. As a result,
we find \beq
    e_\ell(rm) \sim g_\ell \log(rm)
\eeq
where
\beq\label{gl}
    g_{\ell} =
    \frac{ (-1)^{\ell+1} }{
    8 \ell}
     \lt[
\left(%
\begin{array}{c}
 2\ell-2 \\
  \ell-1 \\
\end{array}%
\right)  J_\ell(0)^2 -
 \sum_{j=1}^{\ell}\sum_{k=1}^{j-1} \sum_{q=\pm}
 \left(%
\begin{array}{c}
  2\ell-1 \\
  \ell-j \\
\end{array}%
\right)     (-1)^{j} J_\ell\lt(q(j-k)\rt)^2 \rt]
\eeq
with
\beq
    J_\ell(a) = \left[\prod_{k=1}^{\ell-1} \int_{-\infty}^{\infty} \frac{d x_{k}}{4\pi \cosh\lt(\frc{x_k}2 +\frc{a \pi i}{2\ell}\rt)}\rt]
    \frc1{\cosh\lt(\frc12 \sum_{j=1}^{\ell-1} x_{j}-\frc{a \pi i}{2\ell}\rt) }
\eeq where we used the fact that the integrals factorise into even
and odd-indexed $x$ variables. The function $J_\ell(a)$ can be
evaluated exactly: \beq \label{Jella}
    J_\ell(a) = \frc{2^{-\ell+1}}{(\ell-1)!} \lt\{\ba{ll}
    \csc(a \pi/2) (-1)^{\frc{\ell}2 -1} a \prod_{j=1}^{\frc{\ell}2-1}( a^2-(2j)^2) & \mbox{($\ell$ even)} \z
    \sec(a\pi/2) (-1)^{\frc{\ell-1}2} \prod_{j=1}^{\frc{\ell-1}2}(a^2 - (2j-1)^2) & \mbox{($\ell$ odd)}
    \ea\rt.
\eeq The expression for the number $g_\ell$ can be simplified to
\beq\label{fleur}
    \frc1{8\ell} \lt(2^{3-2\ell} \lt(\ba{c} 2\ell-3 \\
    \ell-2\ea \rt) - \frc{h_\ell}{\pi^2(\ell-1)^2} \rt)
\eeq where \beq
    h_\ell = \lt\{ \ba{ll}
    \lt(\ba{c} 2\ell-2 \\ \ell-1 \ea\rt) \lt(\ba{c} \ell-2 \\ \ell/2-1
    \ea\rt)^{-2} + 2 \sum_{p=0}^{\ell/2-1} \lt(\ba{c} 2\ell-2 \\ \ell-1 + 2p \ea\rt) \lt(\ba{c} \ell-2 \\
    \ell/2-1+ p
    \ea\rt)^{-2} & \mbox{($\ell$ even)} \z
    2 \sum_{p=0}^{\ell/2-3/2} \lt(\ba{c} 2\ell-2 \\ \ell + 2p \ea\rt) \lt(\ba{c} \ell-2 \\
    \ell/2-3/2- p
    \ea\rt)^{-2} & \mbox{($\ell$ odd)}
    \ea \rt.
\eeq The sum over $\ell$ of the first term in (\ref{fleur}) is
readily seen to be $1/4$, whereas the rest was verified to be
consistent with $-1/12$ by summing 500 terms. Hence, we find \beq
    \sum_{\ell=1}^\infty g_\ell = \frc16
\eeq in agreement with the known logarithmic behaviour from CFT.
In appendix \ref{scaldim}, we provide a numerical analysis of the scaling dimension $\Delta_n$ characterising the short-distance behaviour of
the full two-point function of twist fields from our form factor expansion. Agreement with the known CFT dimension
provides an extremely non-trivial check of the validity of this form factor expansion.

\subsection{Exact expression for $V(\kappa)$}\indent \\

\noindent From the results in the previous two sections, it is
possible to derive an exact expression for $V(\kappa)$ appearing
in (\ref{shla}).

First, comparing (\ref{el}) to (\ref{1ps}), we see that if we set
the reflection matrices to 1, we have $2s_{2\ell}(2rm,\kappa) =
e_\ell(2rm)$. The meaning of this is that the replacement $R\to1$
in $\sum_{\ell=1}^\infty s_{2\ell}(2rm,\kappa)$ precisely provides
the leading small-distance behaviour of the entanglement entropy
in the boundary case, which is related to the bulk case by $
S_A^{\rm boundary}(rm) \sim \frc12 S_A^{\rm bulk}(2rm)\; (rm\to0)$
(up to a finite term). The term $s_1(2rm,\kappa)$ in the boundary
case only contributes a finite term as $rm\to0$, and this holds as
well as for all other terms $s_{2\ell+1}(2rm,\kappa)$ for
$\ell=1,2,3,\ldots$.

In order to understand this, consider first the expression for
$s_{2\ell}$ in (\ref{1ps}). The leading small-$t$ behaviour of the
integrals in the square brackets comes from the region of large
rapidities which is not damped by the hyperbolic cosine factors in
the denominator. Integrating $\theta_{2\ell}$ using the
delta-function, we can change variables to the set
$x_i=\h\theta_{i,i+1}$ for $i=1,\ldots,2\ell-2$, and
$\theta_{2\ell-1}$. Then, we have \beq\label{lecas2}
    \theta_i = \sum_{j=i}^{2\ell-2} (-1)^{j-i} x_j + (-1)^{1+i} \theta_{2\ell-1}
\eeq
and in particular,
\beq
    \sum_{i=1}^{2\ell-1}\theta_i = \sum_{j=1}^{\ell-1} x_{2j-1} +\theta_{2\ell-1}, \quad
    \sum_{i=1}^{2\ell-2} \theta_i = \sum_{j=1}^{\ell-1} x_{2j-1} , \quad
    \sum_{i=2}^{2\ell-1} \theta_i = \sum_{j=2}^{\ell-1} x_{2j}. \label{lorignal2}
\eeq
Hence, the hyperbolic cosine factors in the denominators, involving $x_i$'s and the last two
sums above, do not depend on $\theta_{2\ell-1}$, so that it is
the large $|\theta_{2\ell-1}|$ behaviour of the integrand that determines the singularity at small $t$.
On the other hand, $\theta_{2\ell-1}$ is involved in the argument of every $R$-matrix.
Since $R(i\pi/2-\theta)\sim \pm i$ as ${\rm Re}(\theta)\to \mp\infty$ for $\kappa$ finite, and the opposite for
$\kappa=-\infty$, one can see that the product of $R$ matrices goes to 1 as $\theta_{2\ell-1}\to\pm\infty$
for any $\kappa$. This shows that the replacement $R\to1$ gives the leading behaviour as $rm\to0$.

Second, let us consider a similar change of variable in the odd case (\ref{1ps2}), $x_i=\h\theta_{i,i+1}$ for $i=1,\ldots,2\ell-1$,
and $\theta_{2\ell}$. Then,
\beq
    \sum_{i=1}^{2\ell-1} \theta_i = \sum_{j=1}^{\ell} x_{2j-1} - \theta_{2\ell},\quad
    \sum_{i=2}^{2\ell} \theta_i = \sum_{j=1}^{\ell-1} x_{2j} + \theta_{2\ell},
\eeq
so that all variables are involved in the hyperbolic cosine factors in the denominators. Hence, in this case
the limit $t\to0$ can be taken, and all integrals are still convergent.

Then, subtracting this leading behaviour, and using the known small-$rm$ behaviour of the entanglement entropy
in the bulk case, (\ref{shlabu}), we find that the small-$rm$ behaviour in the boundary case is given by (\ref{shla}) with
\beq\label{Vkappa}
    V(\kappa) = \sum_{\ell=1}^{\infty} c_{\ell}(\kappa)
\eeq
with $c_1(\kappa)$ given by (\ref{neg}) and where
\beqa c_{2\ell}(\kappa) &=&  \label{c2l}
    \frac{ \pi^2 (-1)^{\ell} }{2
    \ell}\left[\prod_{k=1}^{2\ell} \int_{-\infty}^{\infty} \frac{d \theta_{k}}{4\pi}\rt]
     \delta(\theta)\lt[
\left(%
\begin{array}{c}
 2\ell-2 \\
  \ell-1 \\
\end{array}%
\right) \frc{\prod\limits_{j=1}^{2\ell}  R\left(\frac{i\pi}{2}-\theta_{j}\right) - 1}{
    \prod\limits_{j=1}^{2\ell}  \cosh\frac{\hat{\theta}_{j j+1}}{2}} \rt. \\ &&  \lt. -
 \sum_{j=1}^{\ell}\sum_{k=1}^{j-1} \sum_{q=\pm}
 \left(%
\begin{array}{c}
  2\ell-1 \\
  \ell-j \\
\end{array}%
\right)     (-1)^{j}  \frc{ \prod\limits_{i=1}^{2\ell} R\left(\frac{i\pi}{2}-(\theta_{i}+ q\frc{j-k}{2\ell}\pi i)\right) - 1}{
  \prod\limits_{i=1}^{2\ell} \cosh\lt(\frac{\h\theta_{i,i+1}}{2} + q\frc{j-k}{2\ell} \pi i\rt)} \rt] \no
\end{eqnarray}
and similarly
\begin{eqnarray}\label{c2lp1}
    c_{2\ell+1}(\kappa) &=& \frac{ \pi^2 (-1)^{\ell} }{2
    \ell+1}\left[\prod_{k=1}^{2\ell+1} \int_{-\infty}^{\infty} \frac{d \theta_{k}}{4\pi}\rt]\delta(\theta) \\ && \times
  \sum_{j=1}^{\ell}\sum_{k=1}^{j} \sum_{q=\pm}
 \left(%
\begin{array}{c}
  2\ell \\
  \ell-j \\
\end{array}%
\right)     (-1)^{j} q \prod\limits_{i=1}^{2\ell+1}  \frc{
 R\left(\frac{i\pi}{2}-(\theta_{i}+ q\frc{j-k+1/2}{2\ell+1}\pi i)\right)}{
  \cosh\lt(\frac{\h\theta_{i,i+1}}{2} + q\frc{j-k+1/2}{2\ell+1} \pi i\rt)}, \no
\end{eqnarray}
both formulae for $\ell=1,2,3,\ldots$.
These generalise $c_2(\kappa)$ given in (\ref{fc}) (recall that $\theta=\sum_{j} \theta_j$). The derivation of (\ref{Vkappa}) is as follows:
\beqa
    S_A^{\rm boundary}(rm) &=& \frc1{12} \log(m\ep) + \frc{U}2 + \sum_{\ell=1}^{\infty} s_{\ell}(2rm,\kappa) \n
        &\sim& \frc1{12} \log(m\ep)  + \frc{U}2+ \sum_{\ell=1}^\infty c_{\ell}(\kappa) +
            \frc12\sum_{\ell=1}^{\infty} e_{\ell}(2rm)  \n      &=& \frc1{12} \log(m\ep) + \frc{U}2 +
            \sum_{\ell=1}^\infty c_\ell(\kappa)
            + \frc12\lt(S_A^{\rm bulk}(2rm)- U - \frc1{6} \log(m\ep)\rt)  \n
        &\sim& -\frc1{12} \log(2r/\ep) +
            \sum_{\ell=1}^{\infty} c_{\ell}(\kappa).
\eeqa Although we have provided arguments suggesting that the
integrals defining $c_{\ell}(\kappa)$ are convergent, in order for
(\ref{Vkappa}) to be a correct representation of $V(\kappa)$, the
infinite sum over $\ell$ should give a finite result. This is much
more subtle, as form factor expansions are expected to provide
convergent series expansion for finite distances, but not
necessarily at zero distance. Exact evaluations of the first few
coefficients $c_\ell(\kappa)$ for some $\kappa$ below, and
extrapolation to higher $\ell$, give strong indications that the
series is indeed convergent; hence (\ref{Vkappa}) is a correct
representation.

\subsection{Exact and approximate re-summations}\indent \\

\noindent All integrals defining the coefficients $c_\ell(\kappa)$
can be evaluated exactly. For instance, a change of variable
$x_j=e^{\theta_j}$ makes all integrands rational functions, and
multiple integrals can bring logarithmic terms, which can all be
integrated. We have been able to evaluate exactly all coefficients
$c_{\ell}(0)$:
\begin{equation}
    c_{2\ell}(0)=\frac{1}{8 \ell  (2\ell-1)},\qquad c_{2\ell+1}(0)=\frac{\pi}{2^{4\ell+1}}\left(
\begin{array}{c}
  2\ell-1 \\
  \ell-1 \\
\end{array}%
\right)^2-\frac{1}{4 (2\ell+1)},
\end{equation}
extrapolating from the exact values at $\ell=1,2,3$ and 4 obtained with the help of Mathematica. Along with $c_1(0)$ from (\ref{neg}), these re-sum to:
\begin{equation}\label{v0}
    V(0)=\frac{\pi-2}{8}+\sum_{\ell=2}^\infty
    c_{\ell}(0)=\log \sqrt{2}.
\end{equation}
>From this, we directly obtain $V(-\infty)$, since
$c_{2\ell}(-\infty) = c_{2\ell}(0)$ and $c_{2\ell+1}(-\infty) =
-c_{2\ell+1}(0)$. This gives \beq\label{Vminft}
    V(-\infty) = \frac{2-\pi}{8}+\sum_{\ell=1}^\infty
   ( c_{2\ell}(0)-c_{2\ell+1}(0))=0.
\eeq We have also computed the values of $c_{\ell}(-1)$ for $\ell$
up to 8 and obtained the following exact expressions,
\begin{eqnarray}
 c_2(-1) &=& \frac{23}{120}=0.191667...,\nonumber \\
 c_3(-1) &=& \frac{247}{12} + \frac{65}{2\pi } - \frac{315\pi
 }{32}= 0.00335195...,\nonumber\\
 c_4(-1) &=& \frac{7771}{5040} - \frac{134}{9\,{\pi }^2}=0.0333052...,\nonumber\\
 c_5(-1) &=& \frac{364737}{40} - \frac{11788}{{\pi }^3} + \frac{420483}{16\pi } - \frac{2787435\pi }{512}=0.000454195...,\nonumber\\
 c_6(-1) &=& \frac{133054637}{800800} + \frac{53609}{30\,{\pi }^4} - \frac{16387}{9\,{\pi }^2}=0.0137526...,\nonumber\\
 c_7(-1) &=& \frac{2756451471}{112} + \frac{10463106}{{\pi }^5} - \frac{413423405}{6\,{\pi }^3} +
  \frac{113700482239}{1152\,\pi } - \frac{35098069395\,\pi
  }{2048}\nonumber\\&=& 0.0000023...,\nonumber\\
 c_8(-1) &=& \frac{1205078334301}{17867850} - \frac{52153432}{105\,{\pi }^6} + \frac{222090851}{180\,{\pi }^4} -
  \frac{176750962}{225\,{\pi }^2}=0.00753035...
\end{eqnarray}
These values already give a rather good approximation to $V(-1)$:
\begin{equation}\label{vm1}
    V(-1)\approx \frac{5}{4}-\frac{3\pi}{8}+ \sum_{\ell=2}^8
    c_{\ell}(-1)= 0.321966...
\end{equation}
which is compatible with the expected value of $ \log \sqrt{2}=
0.346574...$. Therefore, as expected, in the UV limit the entropy
only depends on whether the magnetic field $h$ is finite (free
boundary conditions) or infinite (fixed boundary conditions). An
interpretation of these results for $V(\kappa)$ is provided in
section \ref{discuss}.

\section{Discussion} \label{discuss}

We recall that our main results are 1) the full form factor
expansion for the entanglement entropy in the boundary Ising model
(\ref{boundaryfinal}) for $\kappa \leq 0$, 2) the full form factor
expansion in the bulk case (\ref{bulkfinal}), and 3) an expression
for the short-distance constant $V(\kappa)$ (\ref{Vkappa}) leading
to its exact value (\ref{resVkappa}). We now discuss our results
for the boundary entanglement entropy in the large-distance and
short-distance limits.

\subsection{Infrared behaviour of the boundary entanglement
entropy}\indent\\

\noindent The form factor expansion (\ref{boundaryfinal})
converges rapidly for large distances. Hence we expect the 2- and
4-particle contributions evaluated in section (\ref{twofour}) to
provide a good approximation of the IR behaviour of the
entanglement entropy for sufficiently large distances. Let us
start by studying the behaviour of the correction (\ref{c1}) for
$t=2rm$ large. In this case it is possible to compute the leading
contribution of the correction by expanding the integrand (except
for the exponential term) in the variable $\cosh\theta -1$ around
the value zero. We find,
\begin{equation}
    s_1(t,\kappa\neq {-1,1})=
   - \frac{1}{8}\sqrt{\frac{\pi}{2}}\frac{\kappa +1}{\kappa-1}\frac{e^{-t}}{t \sqrt{t}}+
    {O}(e^{-t}t^{-5/2}),\label{irs1}
\end{equation}
and
\begin{eqnarray}
    s_1(t,1) &=& \frc12 \sqrt{\frc\pi2}\frc{e^{-t}}{\sqrt{t}} + O(e^{-t}t^{-3/2}) \n
   s_1(t,-1) &=&
    \frac{3}{32}\sqrt{\frac{\pi}{2}}\frac{e^{-t}}{t^2 \sqrt{t}}+
    {O}(e^{-t}t^{-7/2}).\label{irs1m1}
\end{eqnarray}
Recall that these hold for all $\kappa\le 1$.
\begin{figure}[h!]
\begin{center}
\includegraphics[width=12cm,height=8cm]{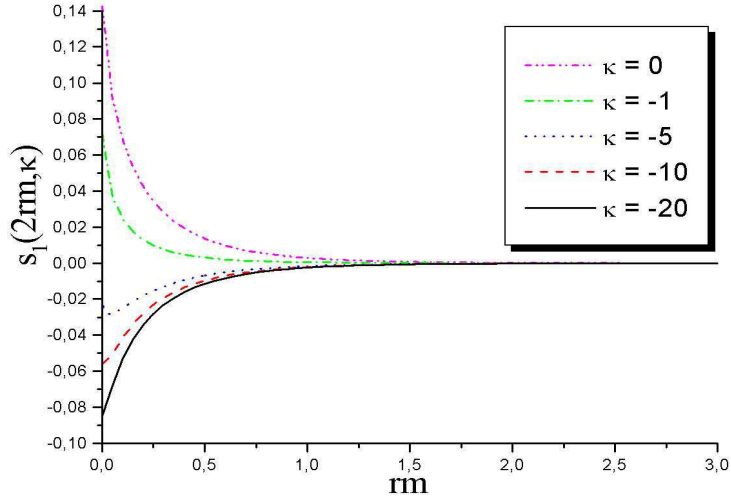}
\end{center} \caption{The function $s_1(2rm,\kappa)$ for several values of $\kappa$.} \label{fig1}
\end{figure}
\begin{figure}[h!]
\begin{flushleft}
\includegraphics[width=7.5cm,height=5.5cm]{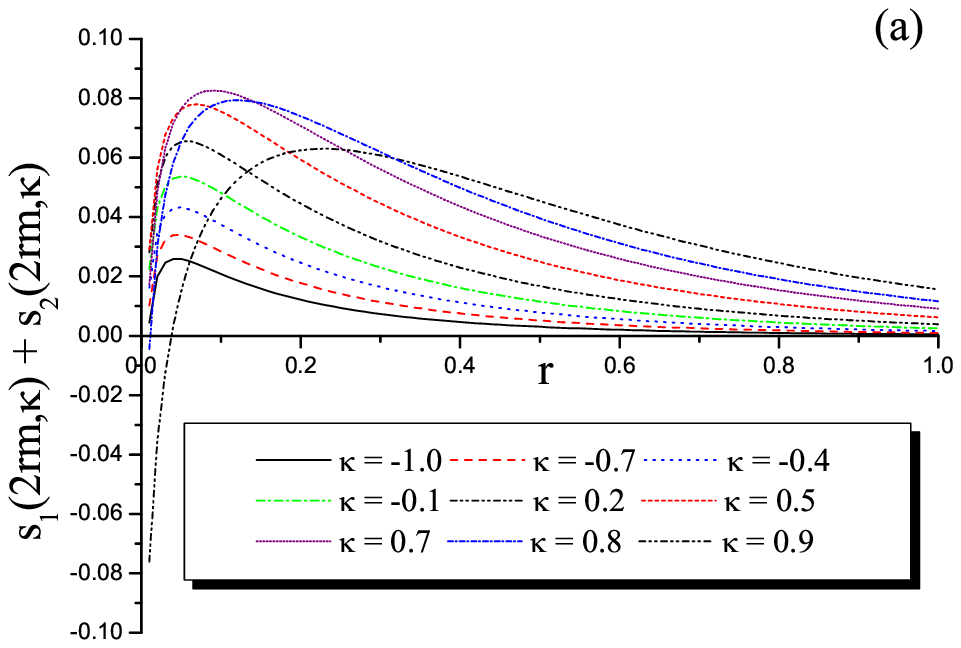}
\includegraphics[width=7.5cm,height=5.5cm]{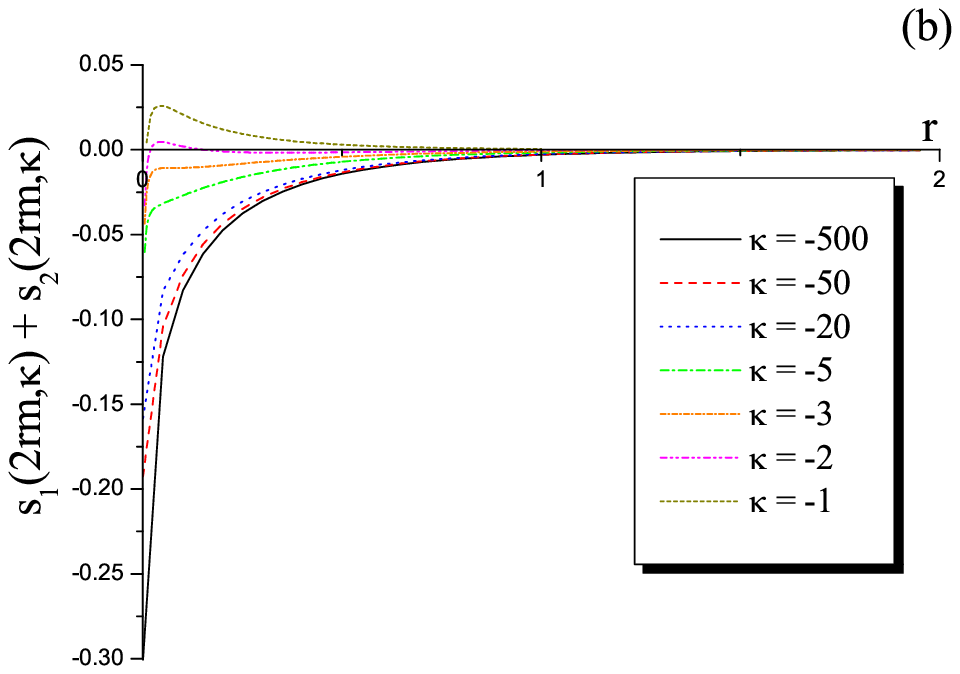}
\end{flushleft}
 \caption{The function $s_1(2rm,\kappa)+s_2(2rm)$ for
$0<r \leq 2$. The value of $rm$ for which the entropy is maximal
for a given value of $\kappa$ increases as $\kappa$ gets closer to
1. As in figures \ref{infra} (a) and (b), the maximum of the
entropy seems most pronounced for $\kappa=0.7$ and $rm$ around 0.1
and tends to shift and eventually disappear as $\kappa \rightarrow
1$. However, since the maximum occurs generally for rather small
values of $rm$, its position may be affected significantly by
higher order form factor corrections. Moreover, for $\kappa>0$, additional contributions
to $s_2$ may come from the boundary bound state. Yet, it is certain that
there is a maximum for $\kappa \geq -1$ as in this case the
function tends to zero from above as $rm \rightarrow \infty$.}
\label{infra}
\end{figure}

A similar computation for $s_2(t,\kappa)$ yields:
\begin{equation}
    s_2(t,\kappa\neq  -1)=-
   \frac{1}{64}\sqrt{\frac{\pi}{2}}\frac{(\kappa +1)^2}{(\kappa-1)^2}\frac{e^{-2t}}{2t \sqrt{2t}}+
   {O}(e^{-2t}(2t)^{-5/2}),
\end{equation}
and
\begin{equation}
   s_2(t,-1)= -
    \frac{15}{1024}\sqrt{\frac{\pi}{2}}\frac{e^{-2t}}{(2t)^3 \sqrt{2t}}+
   {O}(e^{-2t}(2t)^{-9/2}),
\end{equation}
which is clearly sub-leading comparing to the $s_1(t,\kappa)$
expansion. Note that there will be corrections to this sub-leading behaviour for $\kappa>0$ due to the boundary bound state.

Let us examine more carefully (\ref{irs1})
and (\ref{irs1m1}). The leading contribution to the entropy is
negative, as long as $\kappa<-1$: the asymptotic value is approached from below.
For $\kappa> -1$, however, a change of
sign occurs: the asymptotic value is then approached from
above, as can be seen in figure \ref{fig1}. This means that for values of $\kappa > -1$ the ``saturation''
value of the entropy in the infrared limit, $U/2$, is actually not
the maximum value the entropy reaches. For $\kappa
\geq -1$ there exists some finite value of $rm$ for which the entropy has
a maximum. From looking at the curves of $s_1(t,\kappa)+s_2(t,\kappa)$,
this value seems to be moving towards higher $rm$ as $\kappa$ is decreased towards $-1$
-- see figures \ref{infra} (a) and \ref{infra} (b). However, for $-1<\kappa<0$ the maximum is at very small
values of $rm$, and for $\kappa>0$, $s_2$ may receive extra
boundary bound state contributions. Hence, this behaviour of the maximum cannot be conclusive.

The fact that the entanglement entropy is not monotonic for
$\kappa>-1$ is not in contradiction with its fundamental
properties. In particular, the fact that the entanglement entropy
is an increasing function of $rm$ in the bulk case follows from
the ``strong subadditivity theorem" and translation invariance, as
proven in \cite{Casini,Casini:2003ix}. Since translation
invariance is broken in the boundary theory, the entanglement
entropy in this case is not necessarily a monotonic function of
$rm$.

As recalled in sub-section \ref{sectRIsing}, for $\kappa>-1$, the R-matrix
has a pole on the imaginary $\theta$ line, although it is only on the
physical strip for $\kappa>0$ (where a bound state is present).
The value $\kappa=-1$ (corresponding to a magnetic field
$h_c=2\sqrt{m}$) is a ``critical'' value where the $R$-matrix has a
third-order zero at $\theta=0$. What we
see here is that it is this critical value that plays
an important role for seeing a maximum from the large-distance
behaviour. Note that a similar sort of large-distance behaviour
was observed for the one-point functions of the energy and spin operators
in the boundary Ising model \cite{Konik:1995ws}. The explanation suggested was
that while at short distances one sees the free boundary condition for any finite $h$,
at large distances one observes the fixed boundary condition for $h>h_c$,
and the free one for $h<h_c$. This gives rise to a cross-over behaviour in the
$h$--$mr$ plane for $h<h_c$, where we observe two separated regions, corresponding to increasing and decreasing
entanglement entropy with distance.

\subsection{Ultraviolet behaviour of the boundary entanglement
entropy}
\subsubsection{Interpretation of $V(\kappa)$}

It is natural to interpret $V(\kappa)$ as a ``boundary
entanglement'': the contribution of the boundary to the
entanglement between the region $A$ and the rest. Naturally, for
fixed boundary condition, there should be no contribution at all,
since the boundary does not experience quantum fluctuations. Our
result (\ref{resVkappa}) shows that we have chosen the correct
large-distance normalisation to have $V(-\infty)=0$. On the other
hand, for free boundary conditions, the boundary fluctuates and
should participate to the entanglement. This is in agreement with
$V(\kappa>-\infty)=\log\sqrt{2}>0$.

In fact, we may connect $V(\kappa)$ to the {\em boundary entropy} $s$, a
quantity that essentially counts the number of degrees of freedom
pertaining to a boundary. This quantity is simply given by $s=\log g$ where
$g$ is the boundary degeneracy
introduced by Affleck and Ludwig \cite{Affleck:1991tk}. In
particular for a bulk CFT, they showed that $g=\bra0|\t{B}\ket$
where $|0\ket$ is the bulk CFT ground state, and $|\t{B}\ket$ is a
normalised boundary state in the bulk CFT Hilbert space (in particular, $s\leq 0$). It is
considering the boundary entropy that Friedan and Konechny
\cite{Friedan:2003yc} were able to provide a proof of the
``$g$-theorem'': that the $g$-function decreases in the RG flow
from UV to IR.

In order to derive the relation between $V(\kappa)$ and $s$, we first provide a direct connection between $V(\kappa)$ and
entanglement entropies of the Ising model {\em at criticality}:
\beq\label{calca}
    V(\kappa) = S_A^{\rm boundary}(r)_{\rm critical} - \frc12 S_A^{\rm bulk}(2r)_{\rm critical} + \log\sqrt{2}.
\eeq
The result is independent of $r$, since both entanglement
entropies have the same logarithmic $r$-dependence. Most
importantly, both entanglement entropies must be evaluated in the
same cut-off scheme; for instance, both should be evaluated on the
same lattice, with the same lattice spacing. The entanglement entropies should be evaluated at the UV fixed point if $\kappa$ is
finite, and at the IR fixed point if $\kappa=-\infty$.

In quantum field theory, it is not easy to implement the same
lattice spacing in the bulk and boundary situations, since the
lattice spacing enters into non-universal constants. Using massive
QFT, it is possible to solve this problem. Consider the entanglement entropy $S_A(r_1,r_2)$
of sub-section \ref{gencons}, but with a slightly different
normalisation specified below; we will denote it
$\t{S}_A(r_1,r_2)$. We may uniquely fix the cutoff, for instance
by requiring the conformal normalisation as above,
$\t{S}_A(r_1,r_2)\sim \frc{c}3 \log((r_2-r_1)/\varep) + o(1)$ as
$r_2\to r_1$. This, then, is just the bulk critical entanglement
entropy above: $S_A^{\rm bulk}(r)_{\rm critical} = \frc{c}3
\log(r/\varep)$. But using the same object, hence with the same
lattice spacing, we can also define the critical entanglement
entropy in the boundary case, $S_A^{\rm boundary}(r)_{\rm
critical}$, following the arguments already outlined in
sub-section \ref{gencons}. Note first that for $r_2\gg r_1\gg 0$,
the entanglement entropy saturates to some constant $-\frc{c}3
\log(m\varep) + \t{U}$ thanks to the presence of the mass, and
this saturation is a sum of the contributions of the two boundary
points at $r_1$ and $r_2$. These contributions are equal, so that
one boundary point contributes $-\frc{c}6 \log(m\varep) +
\t{U}/2$. Now for $r_2\gg r_1$ and $r_1m$ finite, we get the
entanglement entropy in the boundary case (i.e. with the region
ending on the boundary being connected) and with a mass, but with
an extra contribution of the boundary point $r_2$ (at infinity).
Hence, we may define $\t{S}_A^{\rm boundary}(r) =
\lim_{r_2\to\infty} S(r,r_2) +\frc{c}6 \log(m\varep) - \t{U}/2$.
This quantity has large-$rm$ asymptotic given by $-\frc{c}6
\log(m\varep) + \t{U}/2$. We then only need to take the critical
bulk limit $rm\to0$.

At this point it would seem natural to identify $\t{U}$ with $U$,
the bulk saturation constant, and then have $\t{S}_A^{\rm
boundary}(r) = S_A^{\rm boundary}(r)$. However, this does not
hold. The main idea comes from the fact that in the bulk, the
ground state is degenerate, with spins being asymptotically up or
down, $|0,+\ket$ and $|0,-\ket$. If we allow fluctuations amongst
the two degenerate ground states, as would be obtained from a
large-volume limit of a periodic space, the bulk constant $U$
should be identified with the two contributions from the boundary
points, and an additional pure entropy contribution coming from
this fluctuation, which should be $\log 2$ (for instance, an extra
contribution of $\log 2$ due to zero-modes was present in results
of \cite{peschel}). Hence we must write \beq\label{UtU}
    U = \t{U} + \log 2.
\eeq
In the boundary case with any non-zero
magnetic field, however, there is a definite ground state that is
chosen at infinity, and the derivation above indeed involves only the boundary
point contribution $\t{U}/2$. Hence the large distance limit of
$\t{S}_A^{\rm boundary}(r)$ is $-\frc{c}6 \log(m\varep) +
U/2-\log\sqrt{2}$ so that from (\ref{shla}), $\t{S}_A^{\rm boundary}(r) = S_A^{\rm boundary}(r) -\log\sqrt2$.
Then we find $S_A^{\rm
boundary}(r)_{\rm critical} =\frc{c}3 \log(r/\varep) +
V(\kappa)-\log\sqrt{2}$, which shows (\ref{calca}).

\begin{figure}
\begin{center}
\includegraphics[width=7.9cm,height=3.5cm]{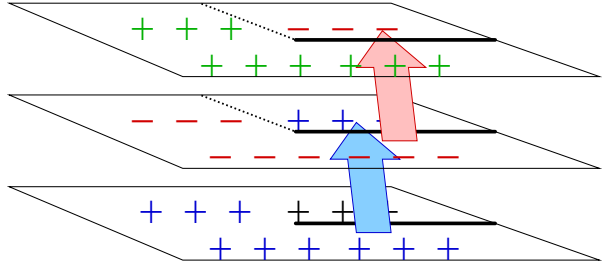}
\end{center}
 \caption{How a vacuum of the form
$\cdots \otimes |0,+\ket \otimes |0,-\ket \otimes |0,+\ket \otimes \cdots$
becomes a two-particle state
after going through a branch-point twist field.}
\label{figtp}
\end{figure}

Technically, allowing fluctuations amongst the two degenerate
ground states corresponds to choosing $|0\ket = ((|0,+\ket +
|0,-\ket)/\sqrt{2})^{\otimes n}$ in the $n$-copy model. Our
branch-point twist field form factors as constructed in
\cite{entropy}, and more generally here, assume the use of this
ground state, since they are non-zero even in cases where there
are sheets with {\em odd numbers} of particles. Recall that in the
ordered regime, particles correspond to domain walls, separating
spin-up and spin-down regions. Then, the only way to have
$out$-states with, for instance, two particles on different sheets
after the twist field, is to have an $in$-state with sheets on
different ground states, see figure \ref{figtp}.
A completely up ground state $|0,+\ket^{\otimes n}$, for
instance, wouldn't admit such form factors. Since all
configurations of an even number of particles distributed amongst the $n$ sheets
are allowed, we must use the completely symmetric ground state.

The constant
$\t{U}$ is the large-distance limit of the bulk entanglement
entropy evaluated from $\bra 0,+|^{\otimes n} \t{\cal T}(r_1)
{\cal T}(r_2) |0,+\ket^{\otimes n}$, whereas $U$ is that obtained
from $\bra 0|\t {\cal T}(r_1) {\cal T}(r_2)|0\ket$ with the
symmetric ground state. The difference can be computed explicitly.
First note that matrix elements
\[
    \bra 0,\ep_1'|\otimes \cdots \otimes \bra 0,\ep_n'| \t {\cal T}(r_1) {\cal T}(r_2) |0,\ep_1\ket\otimes \cdots \otimes |0,\ep_n\ket
\]
have zero large-distance limit unless $\ep_i=\ep_j=\ep_j'$ for all
$i,j$ (that is, all signs are the same), since otherwise domain walls
will have to propagate between the twist fields. Hence we
immediately find
\beq
    \bra 0 |\t {\cal T}(r_1) {\cal T}(r_2)|0\ket \sim 2^{1-n} \bra 0,+|^{\otimes n} \t{\cal T}(r_1) {\cal T}(r_2) |0,+\ket^{\otimes n}
\eeq
at large distances, so that, taking derivatives with respect to $n$, we have
\beq
    -\lt(\frc{d}{dn} \bra 0 |\t{\cal T}(r_1) {\cal T}(r_2)|0\ket \rt)_{n=1} \sim \log 2 - \lt(\frc{d}{dn}
    \bra 0,+|^{\otimes n} \t{\cal T}(r_1) {\cal T}(r_2) |0,+\ket^{\otimes n} \rt)_{n=1}
\eeq
which gives (\ref{UtU}).

Note that although the arguments above hold for finite, and
perhaps only large enough, magnetic field, the constant
$V(\kappa)$ is the same for any $\kappa$ finite (see sub-section
\ref{shortdistbeh}), hence the result (\ref{calca}) is valid in
general\footnote{We would like to thank here P. Calabrese and
P.~E. Dorey, for suggesting to us the possibility of a
relationship between the degeneracy of the ground state and the
$\log \sqrt{2}$ term in (\ref{calca}).}.

Finally, Calabrese and Cardy \cite{Calabrese:2004eu} proposed a formula
for the boundary entropy, which amounts, from (\ref{calca}), to the statement
\begin{equation}
    V(\kappa) = s + \log\sqrt{2},
\end{equation}
and which has been tested numerically in \cite{zhou,igloi} This is
in agreement with our results (\ref{resVkappa}), since $s=0$ in
the free boundary case, and $s=-\log\sqrt{2}$ in the fixed
boundary case \cite{CardyLewellen}.

From these arguments, in general the shift $\log\sqrt{2}$ should
be replaced by $-\log{\cal C}$ where ${\cal C}^2$ is the fraction
of the ground state degeneracy broken by the boundary condition
for large enough $h$. This implies (\ref{resVkappagen}).  Such an
``extra'' contribution to $s=\log g$ was also found in
\cite{Patrick} in calculations using Thermodynamic Bethe Ansatz
techniques and was accounted for by means of similar arguments (it
is important to note that the Thermodynamic Bethe Ansatz used
there is fundamentally different from our approach based on form
factors).

In \cite{Patrick} the
flow of $g$ between critical points was studied for several
families of minimal Toda field theories with ground state degeneracy $k$ and with a boundary completely breaking it.
The factor ${\cal C} = 1/\sqrt{k}$ was termed ``symmetry factor''. Interestingly, it was shown
that at the infrared point of these models, one gets $g={\cal C}$. Hence, for massive models
with spontaneously broken order-parameter symmetry, and with an order-parameter boundary
perturbation, one should find $V(\kappa) = s-s_{IR}\geq 0$, where $s_{IR}=\log{\cal C}$ is the infrared value of $s$.
For the same order-parameter perturbation both on the bulk and boundary, one should simply find $V(\kappa)=s\leq0$.
All these considerations should not depend on integrability.

Note that a more detailed study of $V(\kappa)$ may be useful in
understanding if the boundary entropy $s$ is always bounded from
below.

\subsubsection{Conformal bulk with non-conformal boundary}

Taking the short-distance limit $mr\to0$ while increasing the
magnetic field with a fixed product $2 \kappa mr=a$, we obtain a
conformal bulk theory with a non-conformal boundary, with $a=h^2
r$ the dimensionless parameter relating the distance $r$ to the
boundary and the boundary magnetic field $h$. The entanglement
entropy then interpolates between the UV free-boundary point
$r\to0$, and the IR fixed-boundary point $r\to\infty$. We observed
explicitly this interpolation in the four-particle result in
section \ref{twofour}, through the monotonic function
$c_2^\natural(a)$ (\ref{c2nat}). It is a
complicated matter to generalise this to higher particle
contributions, but it would be interesting to understand if the
function stays monotonic, as this gives a natural $g$-function
flow.

\section{Conclusion}\label{conclu}

The main technical results of this paper are exact infinite-series formulae for the
bi-partite entanglement entropy of the Ising model with and
without boundaries. We have used for this a relationship between
entanglement entropy and the derivative with respect to $n$ at
$n=1$ of correlation functions of branch-point twist fields in the
model composed out of $n$ non-interacting copies of the Ising
model. In order to obtain our formulae, it has been necessary to
tackle several non-trivial intermediate problems:
\begin{itemize}
    \item finding closed expressions for all non-vanishing
    form factors of branch point twist fields in the $n$-copy
    theory,
    \item identifying the correct analytic continuation in $n$ of
    the contributions of these form factors to correlation functions and
     evaluate their derivatives with respect to $n$,
    and
    \item checking both the form factor formulae and their
    analytic continuation for consistency.
\end{itemize}
The first of these problems was easy to solve, as we are dealing
with a free theory for which Wick's theorem applies. Hence, we
have been able to show that all form factors of the twist field
admit expressions in terms of a Pfaffian. Obtaining the right
analytic continuation of every contribution to the form factor
expansion of the twist-field boundary one-point function (or bulk
two-point function) is a highly non-trivial problem. The main
complication arises from the fact that the pole- and
zero-structure of the form factors (in particular, the way some
poles and zeroes cancel each other) changes substantially as soon
as $n$ is allowed to take non-integer values. As a consequence,
the phenomenon observed in \cite{entropy} of non-uniform
convergence of form factors as $n\to1$ is generalised to a
non-uniform convergence as $n\to\ell'$ for all positive integers
$\ell'\leq \ell$ for the $2\ell$-particle form factor. The problem
was solved following the principle that the analytic continuation
is obtained from the analytic function that describes form factor
contributions at values of $n$ large enough. This amounts to
evaluating first the contribution we would obtain analytically
continuing in $n$ around some integer and then adding the residues
of all the extra poles that are crossed by the integration
contours when bringing $n$ from infinity. This is however not the
only analytic continuation that is possible and therefore it
becomes quite crucial to find ways of checking it for consistency.
In this paper we have been able to do this in a very precise
manner by finding an explicit formula for the leading logarithmic
behaviour both of the two-point function of the twist field in the
bulk and of its derivative at $n=1$ (that is, the bulk
entanglement entropy). By a combination of analytical and
numerical computations we have been able to extract both UV
behaviours with extreme accuracy and to show that they agree with
what is expected from CFT arguments.

The results just described have put us in the position to analyse
another quantity of interest in this context, that is the
contribution to the free energy that can be attributed exclusively to
the presence of the boundary. This is essentially the boundary entropy,
the natural logarithm of the boundary degeneracy or $g$-factor
originally introduced by Affleck and Ludwig in
\cite{Affleck:1991tk}. In our analysis we have
computed the universal quantity $V(\kappa)$ which is closely
related to $g$ (where $\kappa$ is related to the boundary magnetic
field in the Ising model). We have defined $V(\kappa)$ as a certain $rm$
independent contribution to the boundary entanglement entropy in
the UV limit. We have found an exact formula for $V(\kappa)$ and
evaluated it exactly at $\kappa=0,-\infty$ to $V(0)=\log\sqrt{2}$
and $V(-\infty)=0$. We have also gathered strong numerical
evidence that $V(\kappa)$ is in fact constant and equal to $V(0)$
for any finite values of $\kappa$. These two values of $V(\kappa)$
would correspond to the two conformal invariant boundary
conditions that are known for the Ising model: the free and fixed
boundary conditions. The fact that $V(\kappa)$ is larger for free
boundary conditions (finite magnetic field) and that
$V(0)-V(-\infty)=\log\sqrt{2}$ are properties which also hold for
$\log g$. However, it is known from Cardy and Lewellen's work
\cite{CardyLewellen} that $g_{\text{free}}=1$ and
$g_{\text{fixed}}=1/\sqrt{2}$, hence
\begin{equation}
    V(\kappa)-\log g=\log\sqrt{2}.
\end{equation}
In this paper we have identified the difference between these two
quantities as an IR contribution to the entanglement entropy coming from the
ground state degeneracy in the periodic case that is broken in the
boundary case. We have proposed a generalisation of this to
more general models with relevant boundary perturbations.

We have also found interesting results for the IR behaviour of the
entanglement entropy: it is known that the bulk entropy saturates
for large distances (in the Ising model, the saturation value is
given by the constant $U$ given in (\ref{u})). In particular, the
fact that the entropy is an increasing function of $rm$ follows
from the ``strong subadditivity theorem" and translation
invariance, as proven in \cite{Casini,Casini:2003ix}. Since
translation invariance is broken in the boundary theory, the
entanglement entropy in this case is not necessarily a monotonic
function of $rm$. Indeed, we find that, for a range of values of
$\kappa$, it has a maximum for some value of $rm$ before reaching
its asymptotic value $U/2$. This range of values of $\kappa$
starts precisely at $\kappa=-1$ which corresponds to a value of
the magnetic field for which, in a sense, the boundary becomes
``critical''.

There are many open problems related to the present work and in
general, to the computation of the entanglement entropy in
integrable QFT. In the case of the bulk Ising model, it is known
that the entanglement entropy can be described via Painlev\'e
transcendents \cite{casini1, casini2}. It would be interesting to
check the consistency of this representation with our full form
factor expansion. In the boundary Ising case, it is known that the
one-point function of the order parameter has a Fredholm
determinant representation for any magnetic field, from which
differential equations can be derived \cite{Konik:1995ws}. It
would be interesting to see if similar formulae hold for the
branch-point twist field. In the general QFT case, the most
obvious problem is perhaps to extend the present analysis to
theories other than the Ising model. We believe that this should
be possible to some extent but it is very unlikely that
re-summations can be done analytically for interacting models. Yet
an independent check of (\ref{resVkappagen}) in the more general
situation would be useful. It would also be interesting to apply
the form factor approach employed here and in
\cite{entropy,other,next} to the computation of the entanglement
entropy of multiply connected regions, both for bulk and boundary
theories and also to extend the analysis of the present paper to
the finite temperature situation.

\paragraph{Acknowledgments:}
We are very grateful to P. Calabrese, J.~L. Cardy and P.~E. Dorey
for discussions and interest in this work. We also would like to
thank the Galileo Galilei Institute for Theoretical Physics in
Florence for hospitality during the program "Low-dimensional
Quantum Field Theories and Applications" where key discussions
occurred. B.D. thanks the Mathematics Group at City University
London and O.C.A. thanks the Mathematics Department of Durham
University, where parts of this work were completed. O.C.A.
acknowledges financial support from a City University London
``Pump Priming" fellowship.

\appendix
\section{Re-summation of fully connected terms} \label{proof}
In this appendix we would like to provide a proof by induction of
the equalities (\ref{for})-(\ref{for2}). Let us then assume that
(\ref{for}) is true for some value of $\ell$ and try to obtain
(\ref{for2}) from it. If (\ref{for}) holds, then it will also hold
when the variable $x_{2\ell}$ is shifted as $x_{2\ell}\rightarrow
x_{2\ell}-2 \pi i p$. We will do this and introduce at the same
time one extra sum in the variable $p$ and one factor $K(y^p)$,
\begin{eqnarray}
&&\sum_{j_1,\ldots,j_{2\ell-1},p=0}^{n-1}K((-x_1)^{j_1})
  K(x_2^{j_1-j_2}) \ldots
  K(x_{2\ell-1}^{j_{2\ell-2}-j_{2\ell-1}})K(x_{2\ell}^{j_{2\ell-1}-p})K(y^p)=
   \frac{(-1)^{\ell} 2 i \sinh(\frac{x}{2})}{\prod\limits_{i=1}^{2 \ell} 2
  \cosh\frac{x_i}{2}}\nonumber\\
  && \qquad\qquad\times \sum_{p=0}^{n-1}\sum_{j=1}^{\ell} \left(%
\begin{array}{c}
  2\ell-1 \\
  \ell-j \\
\end{array}%
\right)
  \left[K(x^{j-p} - i \pi)+K(x^{-j-p} +
  i\pi)\right]K(y^p),\label{above}
\end{eqnarray}
where, as before $x:=\sum_{i=1}^{2\ell}x_i$. We can now employ
(\ref{sum2}) in order to carry out the sum in $p$ in the second
line,
\begin{equation}\label{suminp}
\sum_{p=0}^{n-1}K(x^{\pm j-p} \mp  i\pi)K(y^p)=
\frac{i\cosh\left(\frac{ x+y}{2}\right) }{2\sinh\frac{x}{2}
  \cosh\frac{y}{2}}\left(K(x+y\pm 2\pi i j)+K(x+y\pm 2\pi i
  (j-1))\right).
\end{equation}
Substituting these sums into (\ref{above}) we obtain
\begin{eqnarray}
&&\sum_{j_1,\ldots,j_{2\ell-1},p=0}^{n-1}K(-x_1^{j_1})
  K(x_2^{j_1-j_2}) \ldots
  K(x_{2\ell-1}^{j_{2\ell-2}-j_{2\ell-1}})K(x_{2\ell}^{j_{2\ell-1}-p})K(y^p)=
   \frac{(-1)^{\ell+1} 2\cosh(\frac{x+y}{2})}{2 \cosh\frac{y}{2}\prod\limits_{i=1}^{2 \ell} 2
  \cosh\frac{x_i}{2}}\nonumber \\
  &&\quad \times \sum_{j=1}^{\ell} \left(%
\begin{array}{c}
  2\ell-1 \\
  \ell-j \\
\end{array}%
\right)
  \left[K(x^{j}+y)+K(x^{j-1}+y)+ K(x^{-j}+y)+K(x^{-j+1}+y)\right].\label{above2}
\end{eqnarray}
The sum in $j$ can be split as,
\begin{eqnarray}
&&\sum_{j=1}^{\ell} \left(%
\begin{array}{c}
  2\ell-1 \\
  \ell-j \\
\end{array}%
\right)
  \left[K(x^{j}+y)+ K(x^{-j}+y)\right]+ \sum_{j=0}^{\ell-1} \left(%
\begin{array}{c}
  2\ell-1 \\
  \ell-j-1 \\
\end{array}%
\right)
  \left[K(x^{j}+y)+ K(x^{-j}+y)\right]\nonumber\\
 &&  = \left(%
\begin{array}{c}
  2\ell \\
  \ell \\
\end{array}%
\right) K(x+y) + \sum_{j=1}^{\ell}  \left(%
\begin{array}{c}
  2\ell \\
  \ell-j \\
\end{array}%
\right)\left[K(x^{j}+y)+ K(x^{-j}+y)\right],
\end{eqnarray}
where we have used the identities
\begin{equation}\label{bin}
    2  \left(%
\begin{array}{c}
  2\ell-1 \\
  \ell-1\\
\end{array}%
\right)=  \left(%
\begin{array}{c}
  2\ell \\
  \ell \\
\end{array}%
\right) \qquad \text{and}\qquad  \left(%
\begin{array}{c}
  2\ell-1 \\
  \ell-j \\
\end{array}%
\right) +  \left(%
\begin{array}{c}
  2\ell-1 \\
  \ell-j-1 \\
\end{array}%
\right) = \left(%
\begin{array}{c}
  2\ell \\
  \ell-j \\
\end{array}%
\right).
\end{equation}
This completes our proof, since by calling $y=x_{2\ell+1}$ our
expression (\ref{above2}) is nothing but (\ref{for2}). In an
entirely analogous way it is possible to obtain (\ref{for})
starting with (\ref{for2}).

\section{Alternative formulae} \label{Formulae}

In this section, we re-write formulae (\ref{1ps}), (\ref{1ps2}), (\ref{el}), (\ref{c2l}) and (\ref{c2lp1}). These
are obtained using $\delta(\theta) = \int \frc{d\mu}{2\pi} e^{i\mu\theta}$ and shifting all contours in $\theta$
back to the real line. The results involve less and more symmetric multiple integrals, which is useful for
numerical computations or exact evaluation of the integrals with the help of a symbolic mathematics computer application
(like Maple or Mathematica).

With
\beqa
    A_\ell^e(\mu) &=& \lt(\ba{c} 2\ell-2 \\ \ell-1 \ea\rt) - \sum_{j=1}^{\ell} \sum_{k=1}^{j-1} \sum_{q=\pm} \lt(\ba{c} 2\ell-1 \\ \ell-j\ea\rt) (-1)^j
    e^{\pi \mu q(j-k)} \n
    A_\ell^o(\mu) &=& \sum_{j=1}^\ell \sum_{k=1}^j \sum_{q=\pm} \lt(\ba{c} 2\ell \\ \ell-j \ea\rt) (-1)^j q e^{\pi \mu q(j-k+1/2)},
\eeqa
we have
\beqa
    s_{2\ell}(t,\kappa) &=& \frc{\pi^2 (-1)^\ell}{2\ell} \int_{-\infty}^\infty \frc{d\mu}{2\pi}
    \lt[\prod_{k=1}^{2\ell} \int_{-\infty}^{\infty}
    \frac{d \theta_{k} e^{-t \cosh\theta_k}R\lt(\frc{i\pi}2 - \theta_k\rt)}{4\pi\cosh\frc{\h\theta_{k,k+1}}2} \rt]
    A_\ell^e(\mu) e^{i\mu\theta}
    \n
    s_{2\ell+1}(t,\kappa) &=& \frc{\pi^2 (-1)^\ell}{2\ell+1} \int_{-\infty}^\infty \frc{d\mu}{2\pi}
    \lt[\prod_{k=1}^{2\ell+1} \int_{-\infty}^{\infty}
    \frac{d \theta_{k} e^{-t \cosh\theta_k}R\lt(\frc{i\pi}2 - \theta_k\rt)}{4\pi\cosh\frc{\h\theta_{k,k+1}}2} \rt]
    A_\ell^o(\mu) e^{i\mu\theta} \n
    e_{\ell}(rm) &=& \frc{\pi^2 (-1)^\ell}{\ell} \int_{-\infty}^\infty \frc{d\mu}{2\pi}
    \lt[\prod_{k=1}^{2\ell} \int_{-\infty}^{\infty}
    \frac{d \theta_{k} e^{-rm \cosh\theta_k}}{4\pi\cosh\frc{\h\theta_{k,k+1}}2} \rt]
    A_\ell^e(\mu) e^{i\mu\theta} \n
    c_{2\ell}(\kappa) &=& \frc{\pi^2 (-1)^\ell}{2\ell} \int_{-\infty}^\infty \frc{d\mu}{2\pi}
    \lt[\prod_{k=1}^{2\ell} \int_{-\infty}^{\infty}
    \frac{d \theta_{k} }{4\pi\cosh\frc{\h\theta_{k,k+1}}2} \rt]
    \lt( \prod_{k=1}^{2\ell} R\lt(\frc{i\pi}2 - \theta_k\rt) - 1\rt)
    A_\ell^e(\mu) e^{i\mu\theta}
    \n
    c_{2\ell+1}(\kappa) &=& \frc{\pi^2 (-1)^\ell}{2\ell+1} \int_{-\infty}^\infty \frc{d\mu}{2\pi}
    \lt[\prod_{k=1}^{2\ell+1} \int_{-\infty}^{\infty}
    \frac{d \theta_{k} R\lt(\frc{i\pi}2 - \theta_k\rt)}{4\pi\cosh\frc{\h\theta_{k,k+1}}2} \rt]
    A_\ell^o(\mu) e^{i\mu\theta}.
\eeqa

\section{Scaling dimension from the form factor expansion of the bulk two-point function} \label{scaldim}

The scaling dimension of branch-point twist fields was evaluated
from their  two-particle form factors, along with the known form
factors of the trace of the stress-energy tensor, for arbitrary
$n>1$ in \cite{entropy} using the $\Delta$-sum rule \cite{DSC}.
This gave an exact result (in agreement with the CFT result),
because in free models, the stress-energy tensor does not have
form factors with more than two particles. The scaling dimension
can also be recovered from the two-point function of twist fields
themselves, by analysing its short-distance behaviour. This is not
as convenient, however, since it involves re-summing an infinite
number of terms, which contain more and more integrals (all
form-factor contributions), and usually it is superfluous, since
the $\Delta$-sum rule already gives the scaling dimension. But in
our case, it is cornerstone check, because the correct analytic
continuations in $n$ of higher-particle contributions to the
two-point function of twist fields necessitated a non-trivial
choice of integration contours, a problem which did not occur for
the two-particle contribution.

The most convenient way to evaluate the scaling dimension from the
form factor  expansion of the two-point function in free models is
to take the logarithm, so that we are looking for the coefficient
of the small distance logarithmic divergency. Using the fact that
all form factors can be evaluated using Wick's theorem, the
logarithm of the two-point function is just the sum of all
connected contributions. The analytic continuation in $n$ of
connected form factor contributions was obtained in
(\ref{forgood}). Using the same symmetry arguments as in section
\ref{sectbulk} in order to count the number of connected
contributions of a given particle number, we find \beq
    \log\lt(\frc{\bra0|\t{{\cal T}}(r) {\cal T}(0)|0\ket}{\bra {\cal T}\ket^2}\rt) =
     \sum_{\ell=1}^\infty \frc1{2\ell} \lt[\prod_{j=1}^{2\ell}\int \frc{d\theta_j}{2\pi} e^{-rm\cosh\theta_j}\rt]
    u_{\ell}(\theta_1,\ldots,\theta_{2\ell})
\eeq
where $u_{\ell}(\theta_1,\ldots,\theta_{2\ell})$ is $n$ times the right-hand side of (\ref{forgood}).

As in sub-section \ref{exactlogbound}, it is possible to obtain
the small-$rm$  logarithmic divergency by integrating out the
$\theta_{2\ell}$ variable, after the change of variables to the
set $x_j=\h\theta_{j,j+1},\,j=1,\ldots,2\ell-1$ and
$\theta_{2\ell}$. Separating into even and odd-indexed $x$
variables and using Fourier transform, we can evaluate all
remaining integrals in a similar fashion to what was done in
section \ref{exactlogbound}. In fact, it is convenient to shift
back all contours in (\ref{forgood}) to unshifted contours,
picking up residues. The contribution to
$u_{\ell}(\theta_1,\ldots,\theta_{2\ell})$ of the sum of all
unshifted contours is simply \beq
    \frc{n (-1)^\ell 2i\sinh\theta}{\prod_{j=1}^{2\ell} 2\cosh\frc{\h\theta_{j,j+1}}{2} }
     \sum_{j=1}^\ell \lt(\ba{c} 2\ell-1 \\ \ell-j \ea\rt) F_\ell(\theta)
\eeq
where $\theta=\sum_{j=1}^{2\ell} \theta_j$ and
\beq
 F_\ell(\theta) = K(2\theta + (2j-1)i\pi) + K(2\theta - (2j-1)i\pi).
\eeq
For the coefficient of $-\log rm$ this gives a contribution
\beq\label{cont1}
    \sum_{\ell=1}^\infty \frc{2ni}{(2\pi)^4\ell} \int_{-\infty}^\infty dy F_\ell(y) G_\ell^2(y) \sinh y
\eeq
where
\beq
    G_\ell(y) = \int_{-\infty}^\infty \frc{da}{2\pi} \frc{\pi^\ell e^{iay}}{\cosh^\ell a\pi}.
\eeq
The sum of the residues taken when unshifting the contours gives the following corrections to this contribution:
\beqa
    && \sum_{\ell=1}^{\infty}
    \sum_{j=1}^\ell \sum_{k=1}^{[j/n]} \sum_{q=\pm}{}'  \frc{in(-1)^{k+\ell+1}}{8\pi \ell}
     \lt(\ba{c} 2\ell-1 \\ \ell-j \ea\rt) \sinh \lt( \lt(nk-j+\frc{q+1}2\rt)i\pi\rt)
    \times \n && \times {\rm Re}\lt(J_\ell^2\lt(nk-j+\frc{q+1}2\rt)\rt)
    \label{cont2}
\eeqa where $[\cdot]$ means the integer part, and the function
$J_\ell(a)$ was evaluated in (\ref{Jella}). Here, the sum over $q$
is restricted as follows:
\[
    \sum_{q=\pm}{}' \quad:\quad \mbox{for $k=[j/n]$, the term with $q=+1$ is present if and only if $n[j/n]<j-1$.}
\]
The sum of (\ref{cont1}) and (\ref{cont2}) should be compared with $4\Delta_n = (n-1/n)/12$.

\begin{figure}[h!]
\begin{center}
\includegraphics[width=12cm,height=8cm]{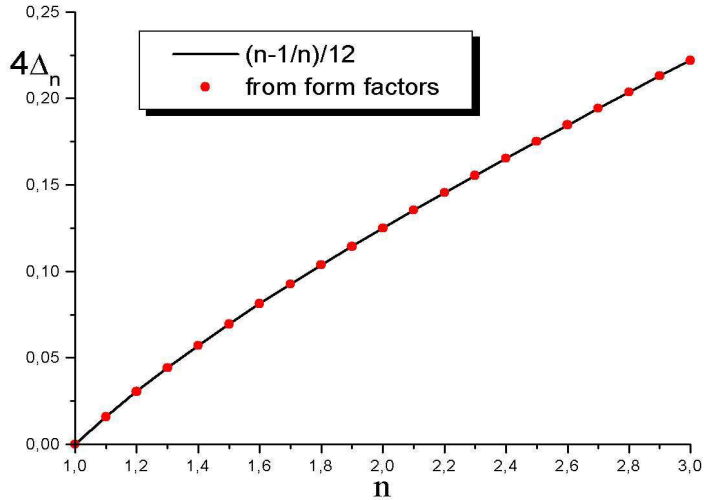}
\end{center} \caption{The exponent in the short-distance behaviour of the bulk
two-point function, as function of the number of sheets $n$
analytically continued to $n\ge 1$. This should be twice the
scaling dimension, and the curve shows the expected value from
CFT.  The data points come from a re-summation of about 30 terms
in the form factor expansion for the logarithm of the two-point
function. For $n>2$, an improvement using Euler's formula is
needed in order to make the series convergent.} \label{figdim}
\end{figure}

Taking about 30 terms in the sums over $\ell$ in both
(\ref{cont1}) and (\ref{cont2}) and adding up both contributions
gives the following numerical values, for $n$ between 1 and 3 (see
fig. \ref{figdim}):
\begin{center}
\begin{tabular}{|l|l|l|l|l|l|l|l|l|l|l|}
 \hline
 $n$ & 1 & 1.1 & 1.2 & 1.3 & 1.4 & 1.5 & 1.6
  \\\hline
  $4\Delta_n\,\,\text{exact}$ & 0 & 0.01591 & 0.03056 &
  0.04423
  &  0.05714 & 0.06944& 0.08125
  \\\hline
  $4 \Delta_n \,\, \text{approx.}$  & 0 & 0.01583 & 0.03047 & 0.04417 &
  0.05710
   & 0.06941 & 0.08123
  \\\hline
\end{tabular}
\end{center}
\begin{center}
\begin{tabular}{|l|l|l|l|l|l|l|l|l|}
 \hline
 $n$ & 1.7 & 1.8 & 1.9 & 2 & 2.1 & 2.2 & 2.3
  \\\hline
  $4\Delta_n\,\,\text{exact}$ & 0.09265 & 0.10370 & 0.11447 &0.125
  &0.13532  &0.14545 & 0.15543
  \\\hline
  $4 \Delta_n \,\, \text{approx.}$  & 0.09264 &0.10370  &0.11447 & 0.12500  & 0.13532&0.14545
  &0.15543
  \\\hline
\end{tabular}
\end{center}
\begin{center}
\begin{tabular}{|l|l|l|l|l|l|l|l|l|}
 \hline
 $n$ & 2.4 & 2.5 & 2.6 & 2.7 & 2.8 & 2.9 & 3
  \\\hline
  $4\Delta_n\,\,\text{exact}$ &0.16528 & 0.175 & 0.18462 &0.19414
  &0.20357  & 0.21293&0.22222
  \\\hline
  $4 \Delta_n \,\, \text{approx.}$  & 0.16530 & 0.17499 & 0.18464& 0.19415 & 0.20358&
  0.21293&0.22192
  \\\hline
\end{tabular}
\end{center}
The corrections (\ref{cont2}) are exactly zero when $n$ is an
integer:  in these cases, no contour shift is necessary from the
beginning. However, the numerical evaluation of (\ref{cont1})
shows that this contribution is insufficient to reproduce the
correct dimension formula for non-integer $n$, showing the
necessity of the contour shifts. It is also interesting to note
that the sum over $\ell$ in (\ref{cont1}) is in fact a divergent
alternating sum for $n>2$. One can however make it convergent by
using Euler's formula for improving convergence of alternating
sums: \beq \sum_{\ell=1}^\infty a_\ell = \sum_{\ell=1}^\infty
2^{-\ell} b_\ell~,\quad b_\ell = \sum_{k=1}^\ell \lt(\ba{cc}
\ell-1 \\ k-1 \ea\rt) a_k. \eeq The rationale behind this is that
the sum is in fact convergent for any $rm>0$,  where Euler's
formula can be used, which then gives a finite limit as $rm\to0$.



\small
\begin{thebibliography}{10}

\bibitem{bennet}
C.~H. Bennet, H.~J. Bernstein, S.~Popescu, and B.~Schumacher,
\newblock Concentrating partial entanglement by local operations,
\newblock Phys. Rev. {\bf A53}, 2046--2052 (1996).

\bibitem{Osterloh}
A.~Osterloh, L.~Amico, G.~Falci, and R.~Fazio,
\newblock Scaling of entanglement close to a quantum phase transition,
\newblock Nature {\bf 416}, 608--610 (2002).

\bibitem{Osborne}
T.~J. Osborne and M.~A. Nielsen,
\newblock Entanglement in a simple quantum phase transition,
\newblock Phys. Rev. {\bf A66}, 032110 (2002).

\bibitem{Barnum}
H.~Barnum, E.~Knill, G.~Ortiz, R.~Somma, and L.~Viola,
\newblock A subsystem-indepent generalization of entanglement,
\newblock Phys. Rev. Lett. {\bf 92}, 107902 (2004).

\bibitem{Verstraete}
F.~Verstraete, M.~A. Martin-Delgado, and J.~I. Cirac,
\newblock Diverging entanglement length in gapped quantum spin systems,
\newblock Phys. Rev. Lett. {\bf 92}, 087201 (2004).

\bibitem{entropy}
J.~L. Cardy, O.~A. Castro-Alvaredo, and B.~Doyon,
\newblock Form factors of branch-point twist fields in quantum integrable
  models and entanglement entropy,
\newblock J. Stat. Phys. {\bf 130}, 129--168 (2007).

\bibitem{other}
O.~A. Castro-Alvaredo and B.~Doyon,
\newblock {Bi-partite entanglement entropy in integrable models with
  backscattering},
\newblock J. Phys. {\bf A41}, 275203 (2008).

\bibitem{next}
B.~Doyon,
\newblock {Bi-partite entanglement entropy in massive two-dimensional quantum
  field theory, 0803.1999}.

\bibitem{bombelli}
L.~Bombelli, R.~K. Koul, J.-H. Lee, and R.~D. Sorkin,
\newblock {A Quantum Source of Entropy for Black Holes},
\newblock Phys. Rev. {\bf D34}, 373--383 (1986).

\bibitem{CallanW94}
J.~Callan, Curtis~G. and F.~Wilczek,
\newblock {On geometric entropy},
\newblock Phys. Lett. {\bf B333}, 55--61 (1994).

\bibitem{Calabrese:2005in}
P.~Calabrese and J.~L. Cardy,
\newblock Evolution of entanglement entropy in one-dimensional Systems,
\newblock J. Stat. Mech. {\bf 0504}, P010 (2005).

\bibitem{Karowski:1978eg}
M.~Karowski,
\newblock {Exact $S$ matrices and form-factors in (1+1)-dimensional field
  theoretic models with soliton behaviour},
\newblock Phys. Rept. {\bf 49}, 229--237 (1979).

\bibitem{ZZ}
A.~B. Zamolodchikov and A.~B. Zamolodchikov,
\newblock Factorized $S$-matrices in two dimensions as the exact solutions of
  certain relativistic quantum field theory models,
\newblock Ann. Phys. {\bf 120}, 253--291 (1979).

\bibitem{abdalla}
E.~Abdalla, M.~C.~B. Abdalla, and K.~D. Rothe,
\newblock Non-perturbative methods in two-dimensional quantum field theory,
  World Scientific, Singapore,
\newblock (1991).

\bibitem{Mussardo:1992uc}
G.~Mussardo,
\newblock {Off critical statistical models: Factorized scattering theories and
  bootstrap program},
\newblock Phys. Rept. {\bf 218}, 215--379 (1992).

\bibitem{Dorey:1996gd}
P.~Dorey,
\newblock {Exact S matrices; hep-th/9810026}.

\bibitem{KW}
M.~Karowski and P.~Weisz,
\newblock Exact S matrices and form-factors in (1+1)-dimensional field
  theoretic models with soliton behavior,
\newblock Nucl. Phys. {\bf B139}, 455--476 (1978).

\bibitem{Smirnovbook}
F.~Smirnov,
\newblock Form factors in completely integrable models of quantum field theory,
\newblock Adv. Series in Math. Phys. {\bf 14}, World Scientific, Singapore
  (1992).

  \bibitem{Cherednik:1985vs}
I.~V. Cherednik,
\newblock Factorizing particles on a half line and root systems,
\newblock Theor. Math. Phys. {\bf 61}, 977--983 (1984).

\bibitem{Sklyanin:1988yz}
E.~K. Sklyanin,
\newblock Boundary conditions for integrable quantum systems,
\newblock J. Phys. {\bf A21}, 2375--2389 (1988).

\bibitem{Fring:1993wt}
A.~Fring and R.~K{\"o}berle,
\newblock Affine Toda field theory in the presence of reflecting boundaries,
\newblock Nucl. Phys. {\bf B419}, 647--664 (1994).

\bibitem{Ghoshal:1993tm}
S.~Ghoshal and A.~B. Zamolodchikov,
\newblock Boundary S matrix and boundary state in two-dimensional integrable
  quantum field theory,
\newblock Int. J. Mod. Phys. {\bf A9}, 3841--3886 (1994).

\bibitem{Fring:1994ci}
A.~Fring and R.~K{\"o}berle,
\newblock Boundary bound states in affine Toda field theory,
\newblock Int. J. Mod. Phys. {\bf A10}, 739--752 (1995).

\bibitem{Bowcock:1995vp}
P.~Bowcock, E.~Corrigan, P.~E. Dorey, and R.~H. Rietdijk,
\newblock Classically integrable boundary conditions for affine Toda field
  theories,
\newblock Nucl. Phys. {\bf B445}, 469--500 (1995).

\bibitem{Affleck:1991tk}
I.~Affleck and A.~W.~W. Ludwig,
\newblock {Universal noninteger 'ground state degeneracy' in critical quantum
  systems},
\newblock Phys. Rev. Lett. {\bf 67}, 161--164 (1991).



\bibitem{CardyLewellen}
J.~L. Cardy and D.~C. Lewellen,
\newblock {Bulk and boundary operators in conformal field theory},
\newblock Phys. Lett. {\bf B259}, 274--278 (1991).

\bibitem{Faddeev:1980zy}
L.~D. Faddeev,
\newblock {Quantum completely integral models of field theory},
\newblock Sov. Sci. Rev. {\bf C1}, 107--155 (1980).

\bibitem{Jin}
B.-Q. Jin and V.~Korepin,
\newblock Quantum spin chain, Toeplitz determinants and Fisher-Hartwig
  conjecture,
\newblock J. Stat. Phys. {\bf 116}, 79--95 (2004).

\bibitem{Konik:1995ws}
R.~Konik, A.~LeClair, and G.~Mussardo,
\newblock On Ising correlation functions with boundary magnetic field,
\newblock Int. J. Mod. Phys. {\bf A11}, 2765--2782 (1996).

\bibitem{casini1}
H.~Casini, C.~D. Fosco, and M.~Huerta,
\newblock {Entanglement and alpha entropies for a massive Dirac field in two
  dimensions},
\newblock J. Stat. Mech. {\bf 0507}, P007 (2005).

\bibitem{casini2}
H.~Casini and M.~Huerta,
\newblock {Entanglement and alpha entropies for a massive scalar field in two
  dimensions},
\newblock J. Stat. Mech. {\bf 0512}, P012 (2005).

\bibitem{Casini}
H.~Casini and M.~Huerta,
\newblock A finite entanglement entropy and the c-theorem,
\newblock Phys. Lett. {\bf B600}, 142--150 (2004).

\bibitem{Casini:2003ix}
H.~Casini,
\newblock {Geometric entropy, area, and strong subadditivity},
\newblock Class. Quant. Grav. {\bf 21}, 2351--2378 (2004).

\bibitem{Friedan:2003yc}
D.~Friedan and A.~Konechny,
\newblock {On the boundary entropy of one-dimensional quantum systems at low
  temperature},
\newblock Phys. Rev. Lett. {\bf 93}, 030402 (2004).

\bibitem{peschel}
I.~Peschel,
\newblock {On the entanglement entropy for a XY spin chain},
\newblock J. Stat. Mech. , P12005 (2004).

\bibitem{Calabrese:2004eu}
P.~Calabrese and J.~L. Cardy,
\newblock Entanglement entropy and quantum field theory,
\newblock J. Stat. Mech. {\bf 0406}, P002 (2004).

\bibitem{zhou}
H.-Q. Zhou, T.~Barthel, J.~O. Fj\ae restad, and U.~Schollw\"ock,
\newblock Entanglement and boundary critical phenomena,
\newblock Phys. Rev. {\bf A74}, 050305(R) (2006).

\bibitem{igloi}
F.~Igl\'oi and Y.-C. Lin,
\newblock Finite-size scaling of the entanglement entropy of the quantum Ising
  chain with homogeneous, periodically modulated and random couplings,
\newblock J. Stat. Mech. , P06004 (2008).

\bibitem{Patrick}
P.~Dorey, A.~Lishman, C.~Rim, and R.~Tateo,
\newblock {Reflection factors and exact g-functions for purely elastic
  scattering theories},
\newblock Nucl. Phys. {\bf B744}, 239--276 (2006).

\bibitem{DSC}
G.~Delfino, P.~Simonetti, and J.~L. Cardy,
\newblock Asymptotic factorisation of form factors in two-dimensional quantum
  field theory,
\newblock Phys. Lett. {\bf B387}, 327--333 (1996).

\end{thebibliography}
\end{document}